\documentclass[a4paper,10pt,3p,twocolumn]{elsarticle}

\usepackage{lineno}
\usepackage{subcaption}
\usepackage{graphicx}
\journal{Astroparticle physics journal}

\begin{document}

\begin{frontmatter}

\title{The Nuclear Window to the Extragalactic Universe}

\author{M.~Erdmann}
\ead{erdmann@physik.rwth-aachen.de}
\author{G.~M\"uller}
\author{M.~Urban}
\author{M.~Wirtz}
\address{Physics Institute 3A, RWTH Aachen University, D-52056 Aachen, Germany}
\begin{abstract}
We investigate two recent parameterizations of the galactic magnetic field 
with respect to their impact on cosmic nuclei traversing the field.
We present a comprehensive study of the size of angular deflections, 
dispersion in the arrival probability distributions, multiplicity 
in the images of arrival on Earth, variance in field transparency, 
and influence of the turbulent field components.
To remain restricted to ballistic deflections, a cosmic nucleus with energy $E$ 
and charge $Z$ should have a rigidity above $E/Z=6$ EV.
In view of the differences resulting from the two field parameterizations as a 
measure of current knowledge in the galactic field, this rigidity threshold may 
have to be increased.
For a point source search with $E/Z\ge 60$ EV, field uncertainties
increase the required signal events for discovery moderately for sources 
in the northern and southern regions, but substantially for sources near 
the galactic disk.
\end{abstract}


\end{frontmatter}


\section{Introduction}

The origin of cosmic rays still remains an unanswered fundamental research question.
Cosmic ray distributions of various aspects have been measured,
most notably the steeply falling spectrum up to the ultra-high energy 
regime with cosmic ray energies even exceeding $E=100$~EeV 
\cite{Abraham:2010mj, AbuZayyad:2013}.

For ultra-high energy cosmic rays, deflections in magnetic fields should diminish
with increasing energy, such that directional correlations should lead to
a straight-forward identification of accelerating sites.
However, even at the highest energies the arrival distributions of cosmic 
rays appear to be rather isotropic.
Only hints for departures from isotropic distributions have been reported, 
e.g., a so-called hot spot \cite{Abbasi2014}, and a dipole signal 
\cite{ThePierreAuger:2014nja}.
At least with the apparent isotropy, limits on the density of extragalactic 
sources were derived which depend on the cosmic ray energy \cite{Abreu:2013kif}.

A recent determination of ultra-high energy cosmic ray composition from 
measurements of the shower depth in the atmosphere revealed contributions of
heavy nuclei above $\sim 5$~EeV \cite{Aab2014a,Aab:2014aea}.
This observation may explain the seemingly isotropic arrival distribution as 
deflections of nuclei in magnetic fields scale with their nuclear charges $Z$.

Obviously, when searching for cosmic ray sources, a key role is therefore 
attributed to magnetic fields.
The galactic field in particular is strong enough to displace original 
arrival directions of protons with energy $E=60$~EeV by several degrees from their
original arrival directions outside the galaxy \cite{Stanev:1996qj}.
The displacement angles for nuclei even reach tens of degrees \cite{Giacinti:2010a}.
The knowledge on the extragalactic magnetic fields is much less certain,
but is likely to be less important than the galactic field \cite{Hackstein:2016pwa}
and is not studied in this contribution.

To identify sources of cosmic rays, rather precise corrections for the propagation 
within the galactic magnetic field are needed which in turn can be used to constrain 
the field \cite{Golup:2009}.
Beyond this, effects of lensing caused by the galactic field have been studied 
which influence the visibility of sources and the number of images appearing 
from a single source \cite{Golup:2011}.
The influence of turbulent contributions to the galactic field has also been 
studied in the context of lensing \cite{Harari:2002} and nuclear deflections 
\cite{Giacinti:2011}.

In previous directional correlation analyses of measured cosmic rays, 
only the overall magnitude of deflections
was taken into account, e.g. \cite{Aartsen:2015dml}, or corrections for cosmic ray 
deflections were applied using analytic magnetic field expressions reflecting the 
spiral structure of our galaxy \cite{Tinyakov:2001ir}.

Recently, parameterizations of the galactic magnetic field have been developed which 
are based on numerous measurements of Faraday rotation \cite{Pshirkov2011,Jansson2012a},
and in addition polarized synchrotron radiation for the second reference.
Based on directional characteristics and the field strength of the parameterizations, 
deflections of cosmic rays 
are predicted to depend strongly on their arrival direction, charge and energy.
In the following we will refer to the regular field with the bisymmetric disk model 
of the first reference as the PT11 field parameterization,
and to the regular field of the latter as the JF12 field parameterization, respectively.

Angular distributions of cosmic rays in these galactic field parameterizations have been
studied before, e.g., with respect to general properties of the JF12 parameterization
\cite{Farrar:2014hma}, specific source candidates \cite{Keivani:2014kua}, 
general properties of deflections and magnifications \cite{Farrar:2015dza, Farrar:inprep}, 
and to the potential of revealing 
correlations between cosmic rays and their sources \cite{emu2015}.

In this work we investigate whether cosmic ray deflections in the galactic magnetic field 
can be reliably corrected for, given the current knowledge of the field.
To simplify discussions of energy and nuclear dependencies we will define rigidity as 
the ratio of the cosmic ray energy and number $Z$ of elementary charges $e$
\begin{equation}
R=\frac{E}{Z\;e} \;.
\end{equation}
In our investigations we use galactic coordinates as our reference system, with
longitude $l$ and latitude $b$.
For a number of visualizations we use Cartesian coordinates alternatively with
height $z$ above the galactic plane, with the Earth being located at 
$(x_E, y_E, z_E)=(-8.5, 0, 0)$~kpc.

Based on the two field parameterizations PT11 and JF12 we initially discuss key
distributions of cosmic ray deflection, dispersion effects in arrival distributions, 
directional variance in field transparency, and the influence of random field components.
From the rigidity dependencies of these distributions, we recommend a minimum
rigidity threshold above which cosmic ray deflection may be controlled in terms
of probability distributions.

Furthermore, we take the different results of the two galactic field parameterizations 
as a measure of our current knowledge of the galactic field.
We compare their cosmic ray angular deflections and study differences in the dispersion of
arrival distributions.
Finally, we study the practical consequences of galactic field corrections and their 
uncertainties by performing simulated point source searches and by quantifying the field
impact in terms of discovery potential.

\section{Field parameterizations}

The two field parameterizations PT11 and JF12 each follow a different ansatz.
Both take into account about $40,000$ Faraday rotation measurements.
The PT11 field has been fitted to two large sets of Faraday rotation measurements.
The JF12 field has been adapted to several large sets of Faraday rotation measurements
and to synchrotron polarization measurements, 
thereby increasing the information per analysed direction by two additional 
complementary measurements \cite{Farrar:2015dza}.
Both use the electron density model NE2001 \cite{cordes2002} with an enlarged 
vertical scale for weighting the line-of-sight integrals of the magnetic field.

\begin{figure}[ttt]
\includegraphics[width=0.45\textwidth]{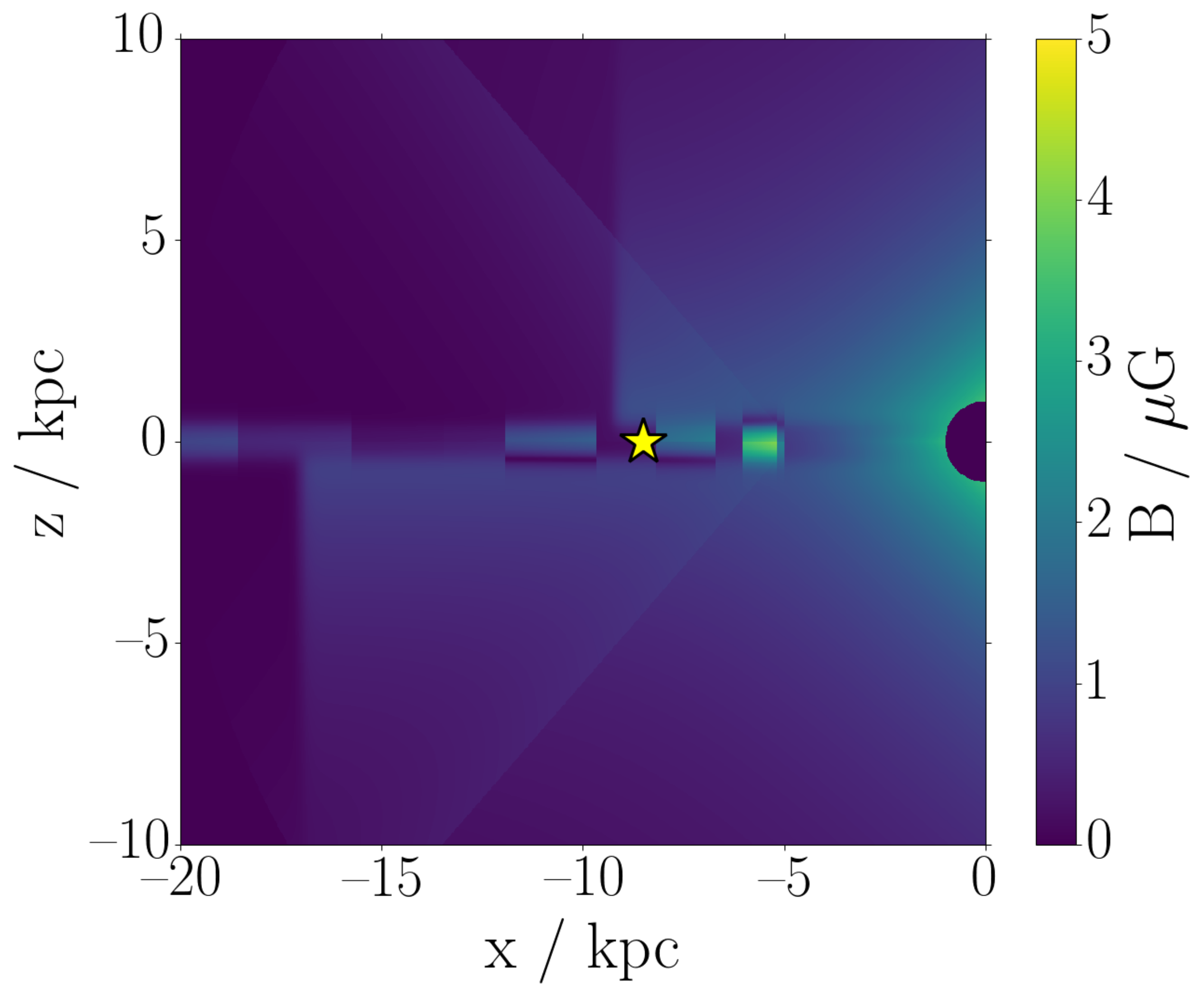}{a)}
\includegraphics[width=0.45\textwidth]{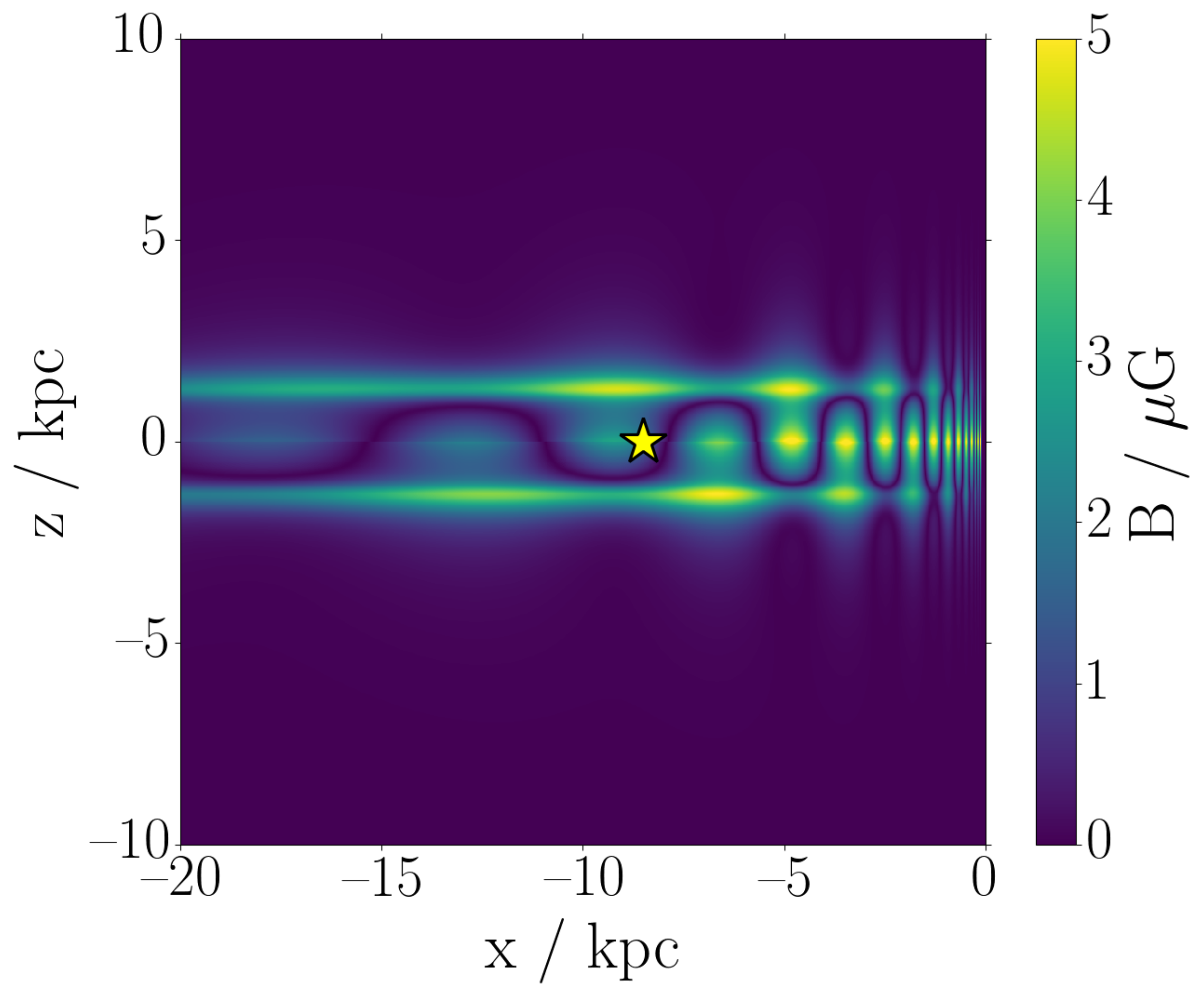}{b)}
\caption{Strength of the galactic magnetic field as a function of the
distance from the galactic center along the solar system line-of-sight, 
and of the distance perpendicular to the 
galactic plane for a) JF12, b) PT11. 
The yellow star denotes our solar environment.}
\label{fig:parameterization}
\end{figure}

Fig.~\ref{fig:parameterization} shows the field strength as a function of the radial 
distance from the galactic center 
along the solar system line-of-sight and the distance perpendicular to the galactic plane.
The fields exhibit different shapes and magnitudes; especially notable in 
Fig.~\ref{fig:parameterization}a is the field extent of the JF12 parametrization above and 
below the galactic plane with non-negligible field strengths even at a distance of
$10$~kpc. 
The PT11 field (Fig.~\ref{fig:parameterization}b), on the other hand, exhibits a rather concentrated 
halo field, which is centered around a distance of $\sim 1.2$~kpc to the galactic plane.

\begin{figure}[hbt!]
\includegraphics[width=0.5\textwidth]{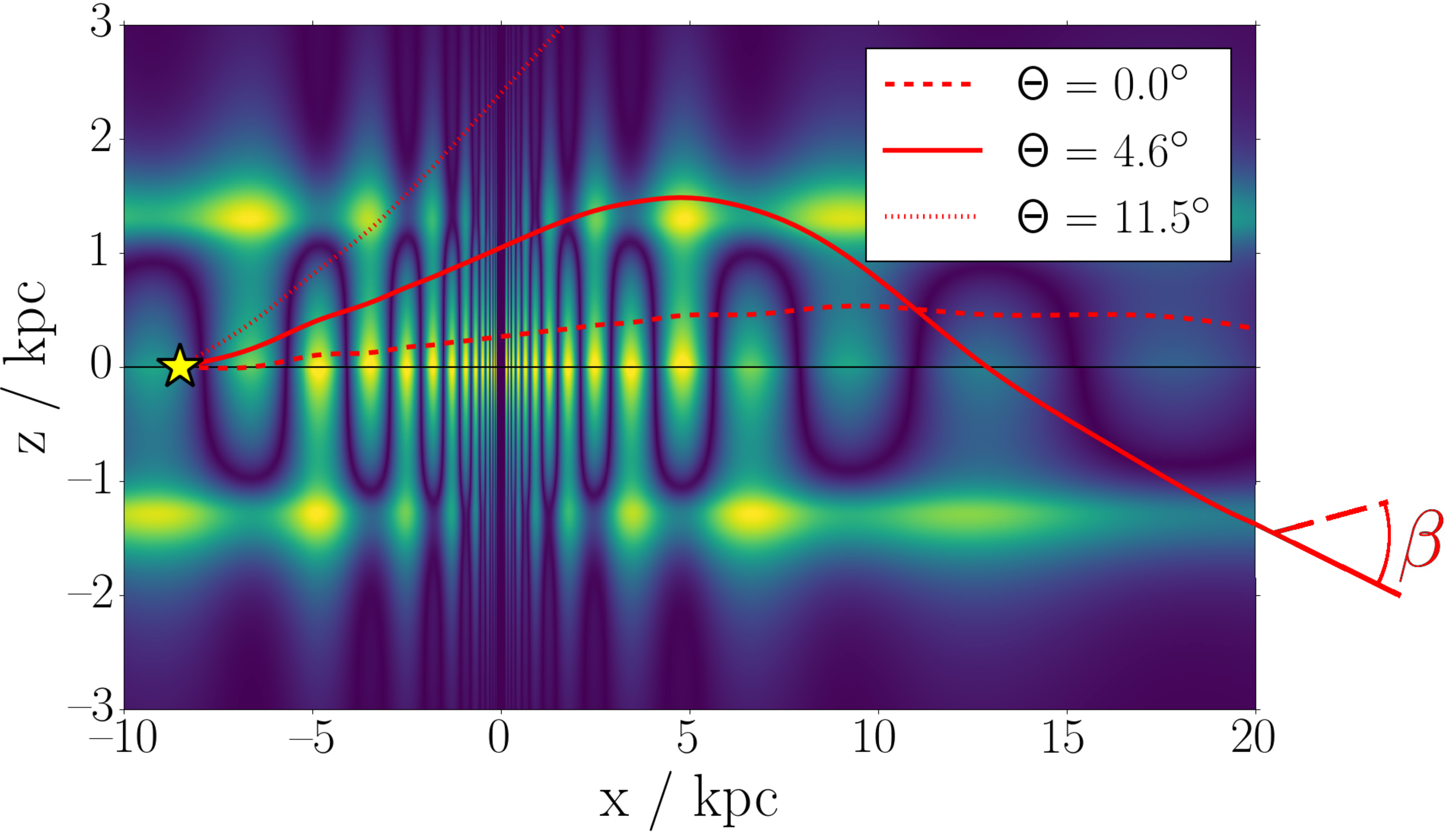}
\caption{Example trajectories of antiprotons originating on Earth
with different initial angular directions $\Theta$ with respect to the galactic plane;
directional change $\beta$ between the direction on Earth 
and the direction outside the galaxy (PT11).}
\label{fig:beta}
\end{figure}

When studying the magnitude of angular deflections of cosmic rays resulting from 
these parameterizations, we take the angle $\beta$ between the 
incoming direction to the galaxy and the arrival direction on Earth 
as a measure of the directional change (Fig. \ref{fig:beta}).

To get a first impression of the different deflections resulting from the two
field parameterizations we use backward tracking techniques of antiprotons 
through the galactic field.
With this technique we obtain individual trajectories for matter particles 
entering from outside the galaxy and then following the reverse path. 
The method ensures that every trajectory leads to observation on Earth.

\begin{figure}[hbt!]
\includegraphics[width=0.45\textwidth]{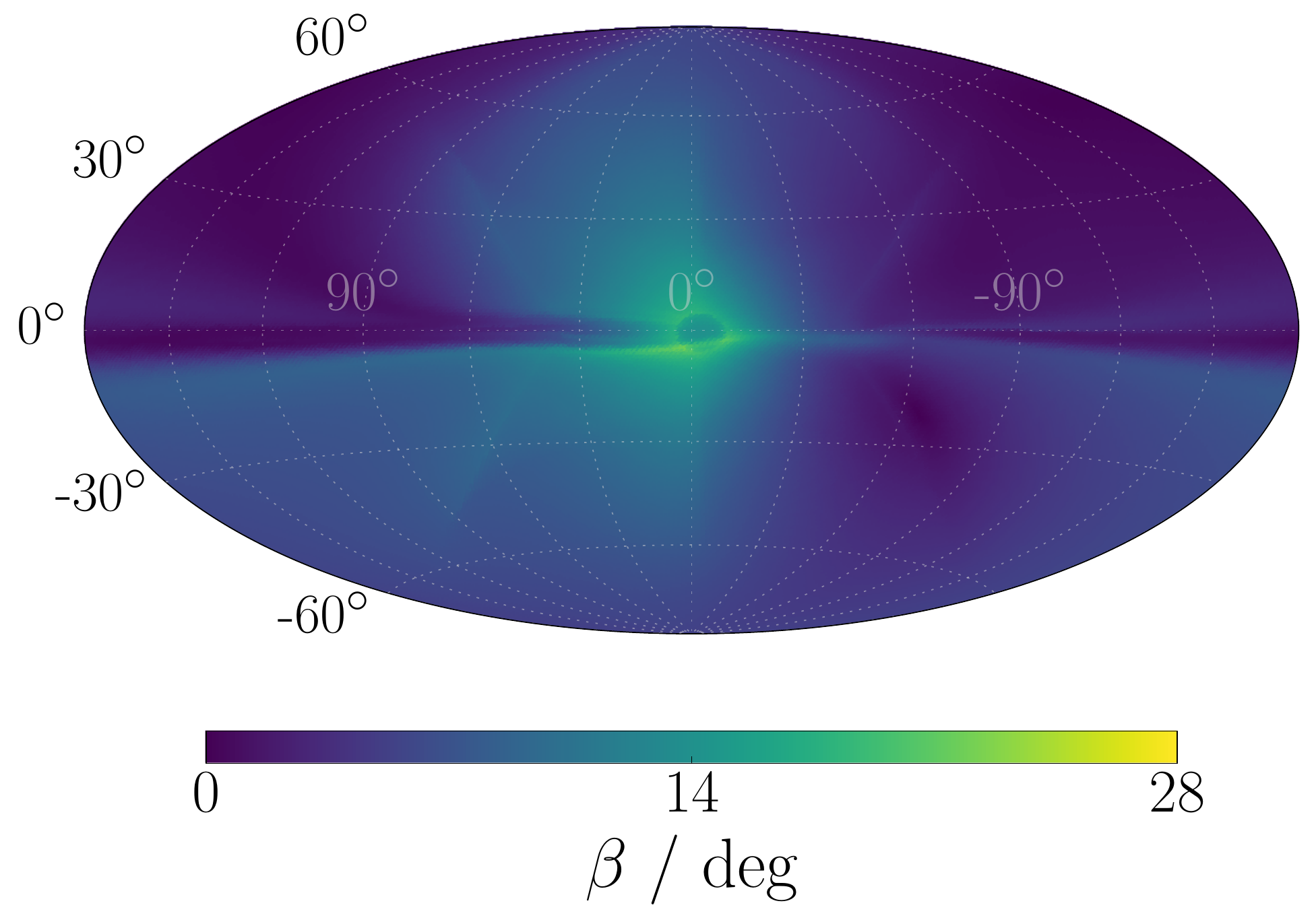}{a)}
\includegraphics[width=0.45\textwidth]{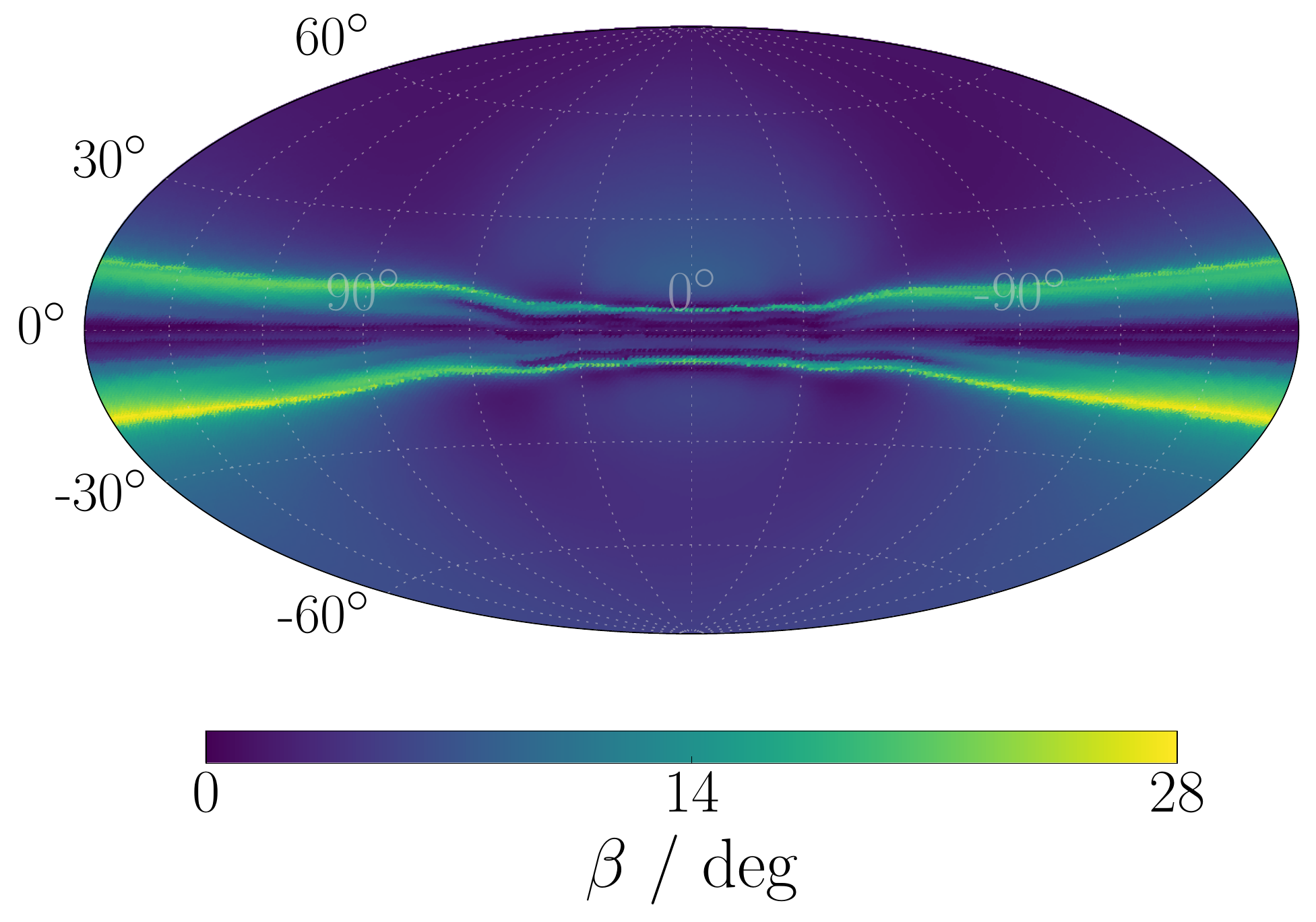}{b)}
\caption{Magnitude of the angular deflections $\beta$ in the galactic magnetic field 
for antimatter with rigidity $R=60$~EV starting from Earth in the direction 
presented in galactic coordinates, 
a) JF12, b) PT11.
}
\label{fig:angularsky}
\end{figure}

In Fig.~\ref{fig:angularsky} we show the magnitudes of the angular deflections $\beta$
of cosmic rays with rigidity $R=60$~EV.
The position in the map denotes the initial direction on Earth in galactic 
coordinates for the backtracked antiprotons. 
The color code refers to the magnitude of angular deflections which reach up 
to $\beta=28$ deg. 

For the JF12 parameterization (Fig.~\ref{fig:angularsky}a), deflections are largest
near directions of the galactic center which is expected from the magnitude of 
the field shown in Fig.~\ref{fig:parameterization}a.
With the PT11 parameterization (Fig.~\ref{fig:angularsky}b), deflections are largest 
in any direction near the galactic plane which is attributed to the strong disk field
(Fig.~\ref{fig:parameterization}b).

As expected, the differences in the field parameterizations relate directly
to a different impact on cosmic ray deflections.
In the following section we study a number of aspects related to the 
directional changes of cosmic rays when traversing the galactic field.

\section{Impact of the galactic magnetic field on cosmic ray arrival}

The goal of this section is to determine a kinematic regime 
where information on cosmic ray arrival directions can be obtained 
by transformation of probability distributions.
For this purpose we first analyze cosmic ray angular deflections as a function of rigidity.
As our primary criterion we require angular deflections to be below $90$~deg 
in order to distinguish ballistic deflections from diffusive-type random walk.

Beyond this we investigate the dispersion of arrival probability distributions 
by the galactic field, and the splitting of arrival distributions into several images.
Furthermore we show directional dependencies of the field transparency for cosmic 
matter and antimatter particles.
We also study the influence of random components of the field which imposes 
uncertainties on the arrival directions.

\subsection{Deflection angles}

The two magnetic field parameterizations exhibit different field strengths
above and below the galactic plane, and differ substantially in their field
characteristics near the plane (Fig.\ref{fig:parameterization}).
Therefore, we divide the sky into three regions of equal 
solid angles, and study angular deflections for each region separately.
The boundaries of these regions are fixed at galactic latitudes of $\pm 19.5$~deg.

To ensure that every cosmic ray trajectory leads to observation on Earth we use the
backward tracking method explained in the previous section.
We use the term ``northern region'' to refer to antiparticles originating on Earth 
in the direction of positive latitudes above $19.5$~deg.
Negative latitudes below $-19.5$~deg, on the other hand, are referred to as the ``southern region'', while for latitudes in-between we use the term ``disk region''.

\begin{figure*}[hht*]
  \begin{minipage}{\textwidth}\footnotesize
  \centering
  \includegraphics[width=0.45\textwidth]{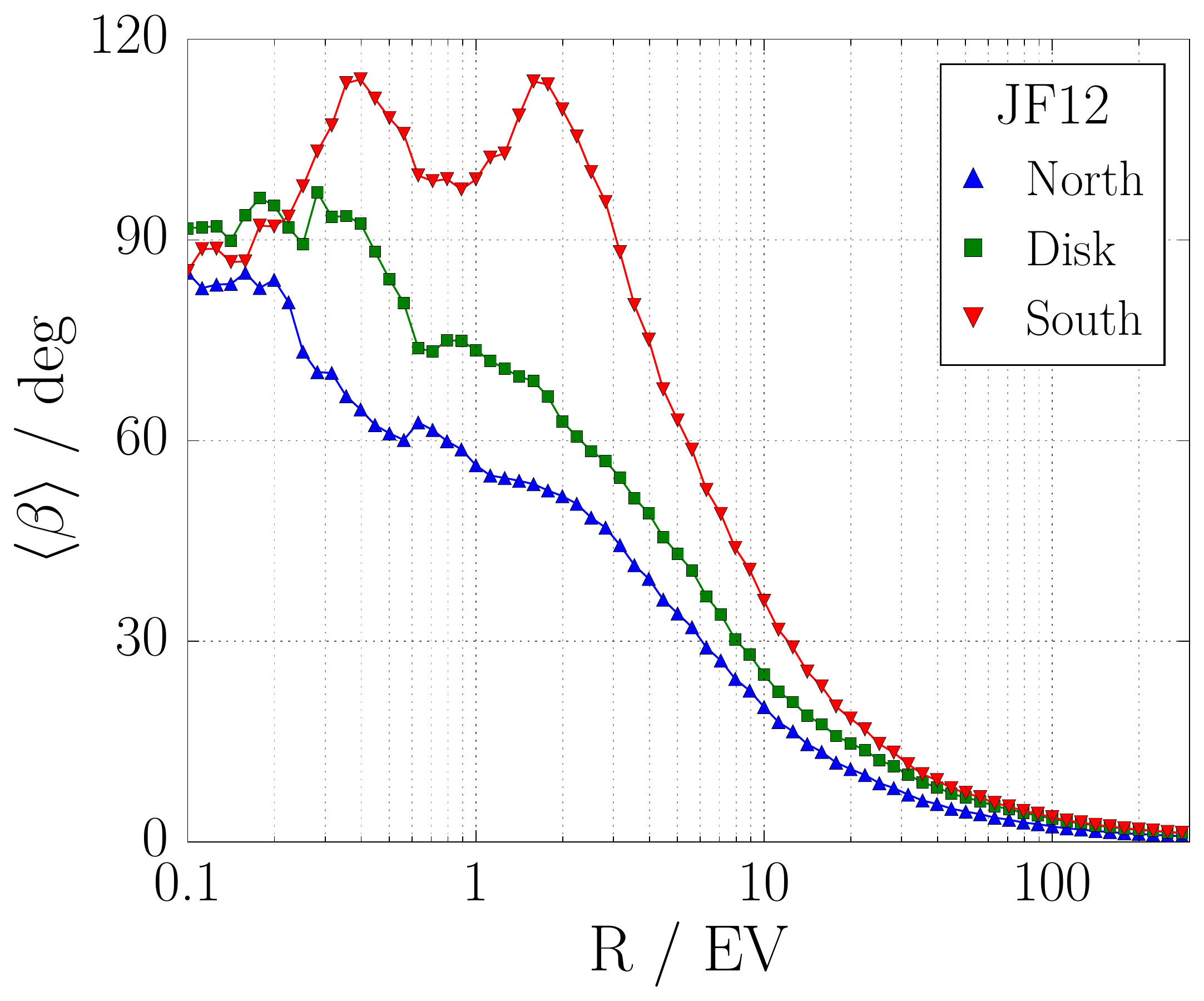}{a)}
  \qquad
  \includegraphics[width=0.45\textwidth]{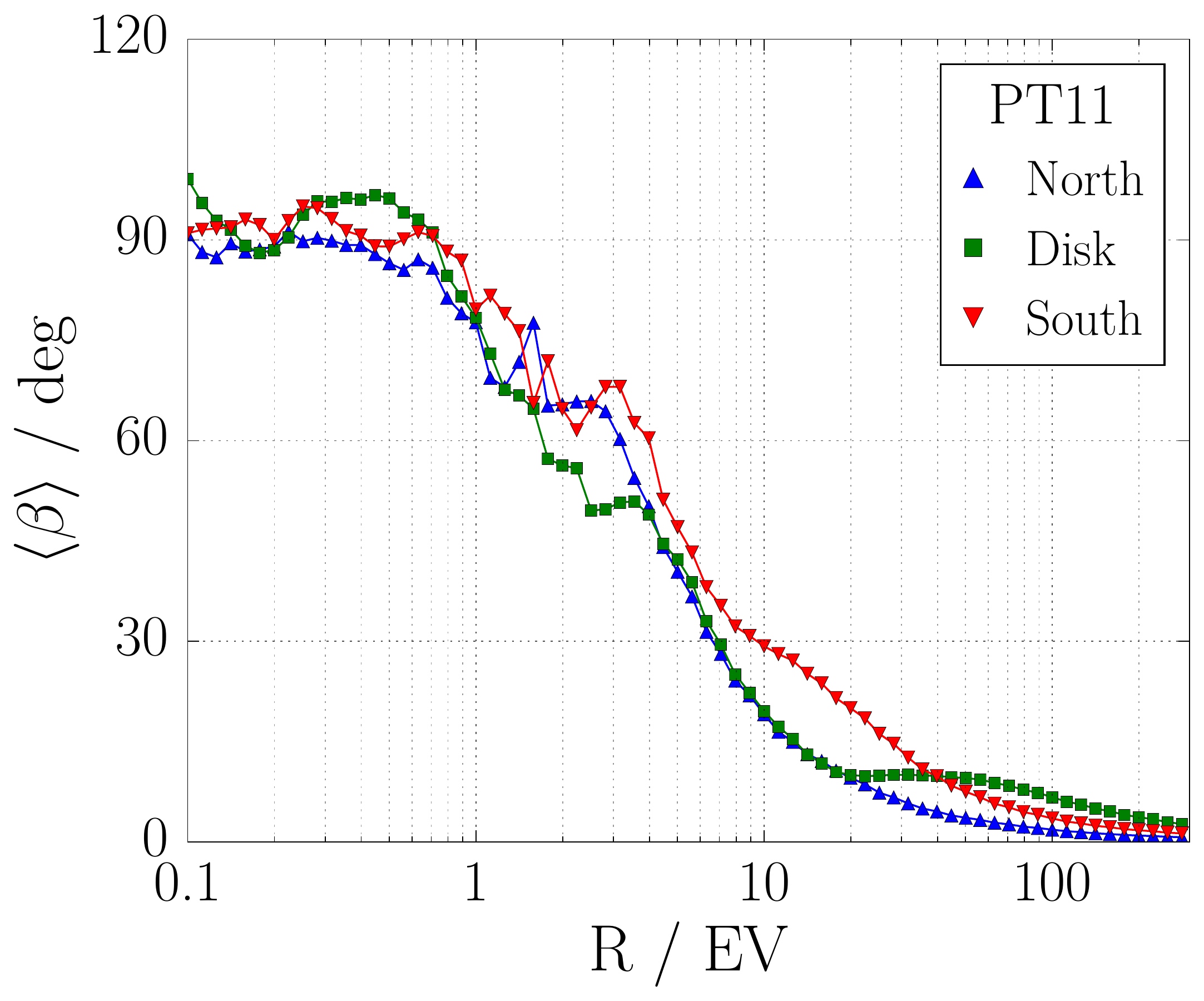}{b)}

  \medskip

  \centering
  \includegraphics[width=0.45\textwidth]{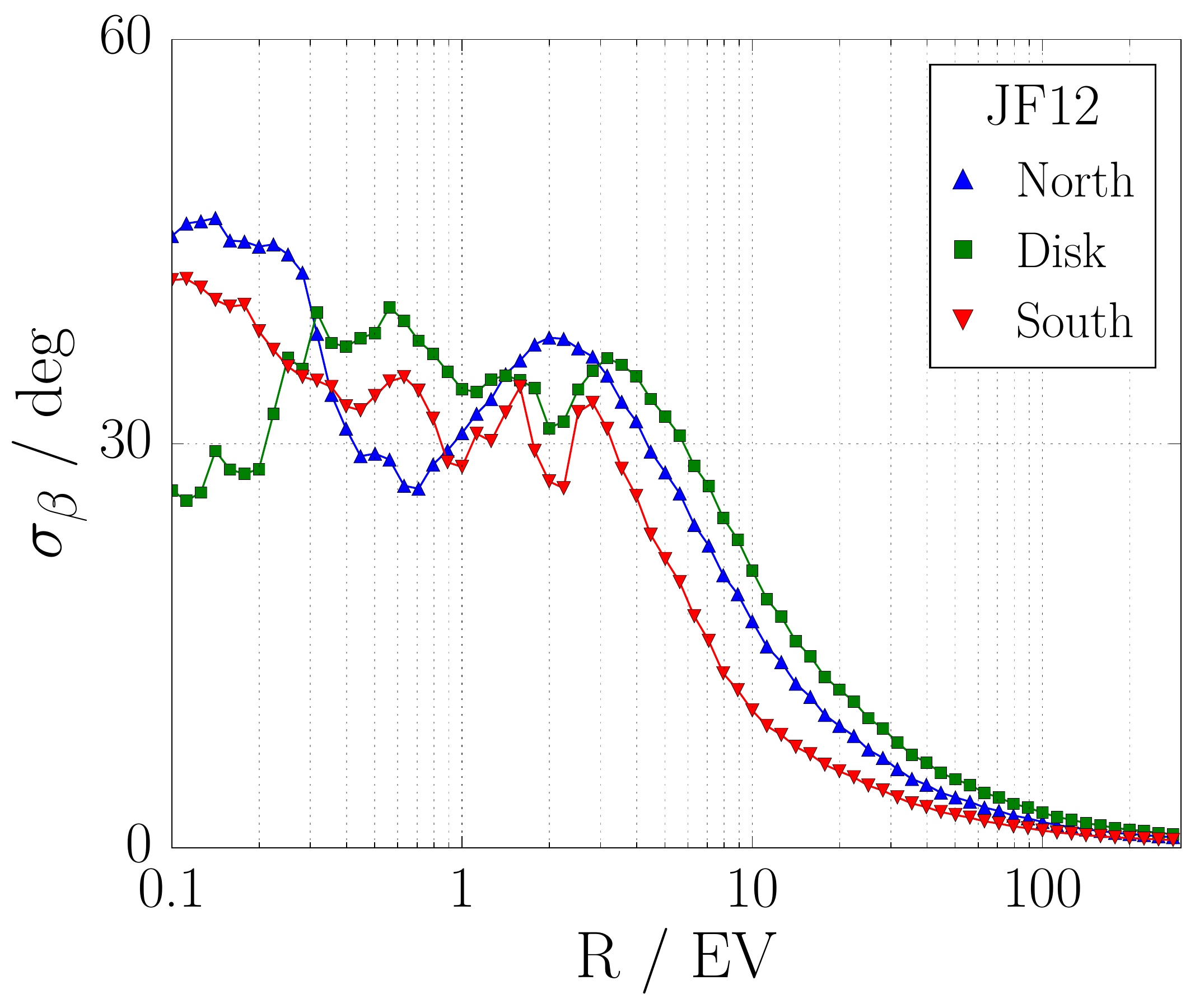}{c)}%
  \qquad
  \includegraphics[width=0.45\textwidth]{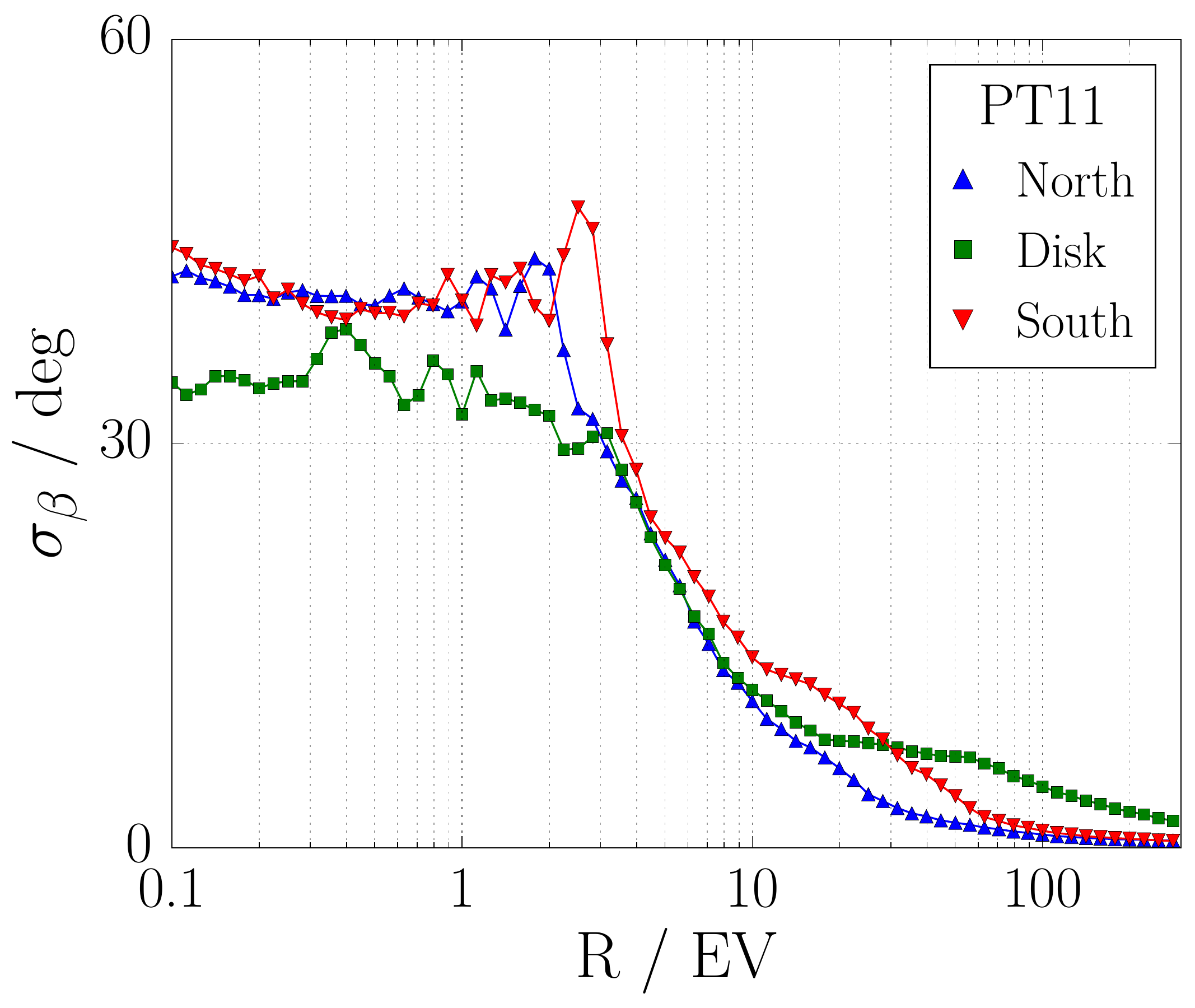}{d)}
  \end{minipage}
\caption{a,b) Average angular deflections $\langle\beta\rangle$
in the galactic magnetic field for antimatter 
originating on Earth in three regions separated by galactic latitudes $\pm 19.5$~deg
as a function of rigidity $R$, 
a) JF12, b) PT11.
c,d) Spread $\sigma_\beta$ of the angular deflections in terms of standard deviations,
c) JF12, d) PT11.
}
\label{fig:angular}
\end{figure*}

In Fig.~\ref{fig:angular}a we show the average directional change 
$\langle\beta\rangle$ between the direction on Earth and the direction 
outside the galaxy as a function of the cosmic ray rigidity $R$ using the JF12 field.
The distribution was derived from $5$ million simulated cosmic rays per 
rigidity interval.
In Fig.~\ref{fig:angular}c we show the corresponding spread $\sigma_\beta$ 
in terms of standard deviations.
For low rigidities $R\sim 0.1$~EV, 
cosmic ray confinement owing to the size of our galaxy and its magnetic field
leads to large directional changes 
$\langle\beta\rangle\sim 90$~deg and large average variations in $\beta$ 
($\sigma_\beta\sim 40$~deg).

At rigidity $R=6$~EV, the largest average deflection of $\langle\beta\rangle\approx 50$~deg 
is found in the southern region (Fig.~\ref{fig:angular}a, downward-pointing triangles).
The corresponding spread amounts to $\sigma_\beta\approx 15$~deg (Fig.~\ref{fig:angular}c), such that
for $95\%$ of the cosmic rays the deflection angle remains below $\beta=90$~deg.
The average deflection in the northern region 
is substantially smaller with only $\langle\beta\rangle\approx 30$~deg, however, 
the spread of $\sigma_\beta\approx 25$~deg is larger (upward-pointing triangles).
Also here most of the cosmic ray deflections are below $\beta=90$~deg.
A similar conclusion holds for the disk region (square symbols).

The PT11 parameterization exhibits very similar tendencies for rigidity $R=6$~EV
as can be seen in Figs.~\ref{fig:angular}b,d.
In the disk region the deflections exceed those of the other regions
for rigidities above $R=40$~EV (square symbols).
Here the deflections also exceed that of the JF12 parameterization as was already visualized 
in Fig.~\ref{fig:angularsky} above.

Overall, at rigidities $R>6$~EV the deflections are consistently reduced and correspondingly enhance the control over cosmic ray deflection.
In the following studies we will therefore use the rigidity of $R=6$~EV
as a benchmark.

\subsection{Dispersion \label{sec:dispersion}}

Cosmic rays originating from a point source may arrive slightly dispersed after 
their propagation through extragalactic fields.
When traversing the galactic field the extent of the arrival distribution may even be  
enlarged.

In order to obtain information on the arrival direction and arrival probability on Earth 
of a cosmic ray that enters the galaxy in any direction, we use a lensing technique 
\citep{Harari2000, Bretz2014}.
The lenses consist of matrices based on the HEALPix format \citep{Gorski2005}, where we 
divide the sphere into $N_{pix}=49,152$ equally sized pixels of approximately $1\;\deg$ in size.
For each rigidity interval a separate matrix is produced by backtracking a set of $N_{pix}\times 100$ 
antiparticles, which are distributed uniformly in each pixel.
The matrices thus contain the probability of a cosmic ray entering the galaxy
with rigidity $R$ at pixel direction $(l_i, b_i)$ to be observed in pixel direction 
$(l^\prime_j, b^\prime_j)$.
As defined above, $l, l^\prime$ refer to the galactic longitudes, and $b, b^\prime$ to the
latitudes, respectively.

By design, the lenses project an extragalactic isotropic distribution onto 
an isotropic distribution on Earth.
Note that some incoming directions have more simulated trajectories leading to observation 
on Earth, while other directions have less, such that there are directionally dependent variations in the transparency of the field.
We normalize the lenses to ensure that the lenses return relative arrival probabilities,
and that an isotropic cosmic ray flux is preserved.
However, for cosmic rays arriving from individual sources the flux varies depending on the source directions which we will show below.
The technical details of the lenses and their production are outlined in \citep{Bretz2014, winchen2013}.
The lenses used in this contribution were calculated with the CRPropa v$3$ program 
\citep{Batista2016} using the PT11 and JF12 parameterizations.

\begin{figure}[h]
\centering
\includegraphics[width=0.45\textwidth]{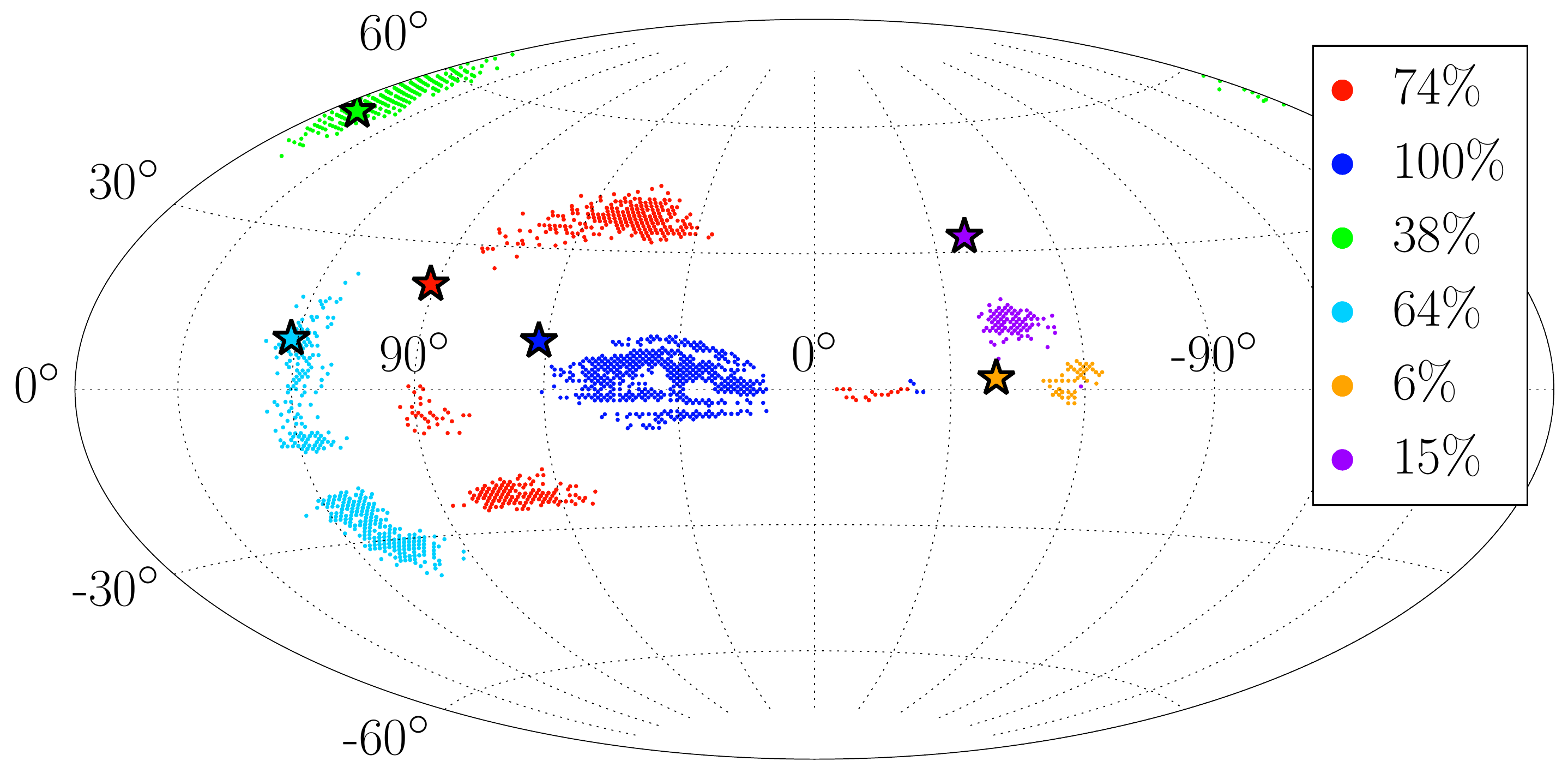}
\caption{Arrival distributions for cosmic rays with rigidity $R=10$~EV originating
from arbitrarily chosen sources (star symbols), and the relative arrival probability in percent.
Relations of cosmic rays with their sources and corresponding arrival probabilities are 
indicated by the color code (JF12).}
\label{fig:images}
\end{figure}

In Fig.~\ref{fig:images} we show examples of arrival directions on Earth 
for simulated cosmic rays with rigidity $R=10$~EV together with their 
original source directions using the JF12 parameterization of the regular field.
The incoming cosmic rays followed a Fisher probability distribution \citep{Fisher1953}
$f(\alpha, \kappa)=\kappa \exp{(\kappa \cos{\alpha})} / (4\pi\sinh{\kappa})$
with a Gaussian width of $\sigma=1/\sqrt{\kappa}=3$~deg.
Their arrival directions were calculated with the 
lensing techniques described above.
The color code indicates the relation between sources and their cosmic rays.
For all scenarios, the fraction of arriving cosmic rays is shown on the right side of the 
figure, normalized to the source with the highest arrival probability.

Different images of the cosmic rays appear depending on their incoming direction.
For example, a source direction which coincides on average with the cosmic rays after 
traversing the galactic field is denoted by the green symbols.
Only a widening of the directional distribution is observed.
Another example is a source direction where the cosmic ray distribution is displaced 
without a strong spread (purple symbols).

Examples of source directions where the cosmic rays are substantially deflected and 
exhibit a wide-spread distribution of arrival directions are denoted by the
light blue and dark blue symbols.
For some source directions, small variations in the cosmic ray incoming direction 
lead to largely different paths, and therefore to several distinct images of the
arrival directions (red symbols).

To investigate dispersion effects in the galactic field we again choose initial 
cosmic rays to follow a Fisher probability distribution with a Gaussian width of $3$~deg.
In principle this value could be related to dispersion effects caused by extragalactic 
fields, which implies a dependency on cosmic ray rigidity.
However, to ensure clarity of our galactic field investigations we will use a fixed 
Gaussian spread for all cosmic rays incoming to our galaxy throughout this work.

After the cosmic rays traversed the galactic field we quantify the direction and extent of 
the resulting arrival distribution by calculating around every HEALPix pixel 
$(l^\prime_j, b^\prime_j)$ with non-zero probability a circular curve which includes
$50\%$ of all arrival probabilities.
We then use the radius $r_{50}$ of the smallest circle as a measure of the extent 
of the probability distribution \citep{gmueller2016}.

\begin{figure}[h]
\includegraphics[width=0.45\textwidth]{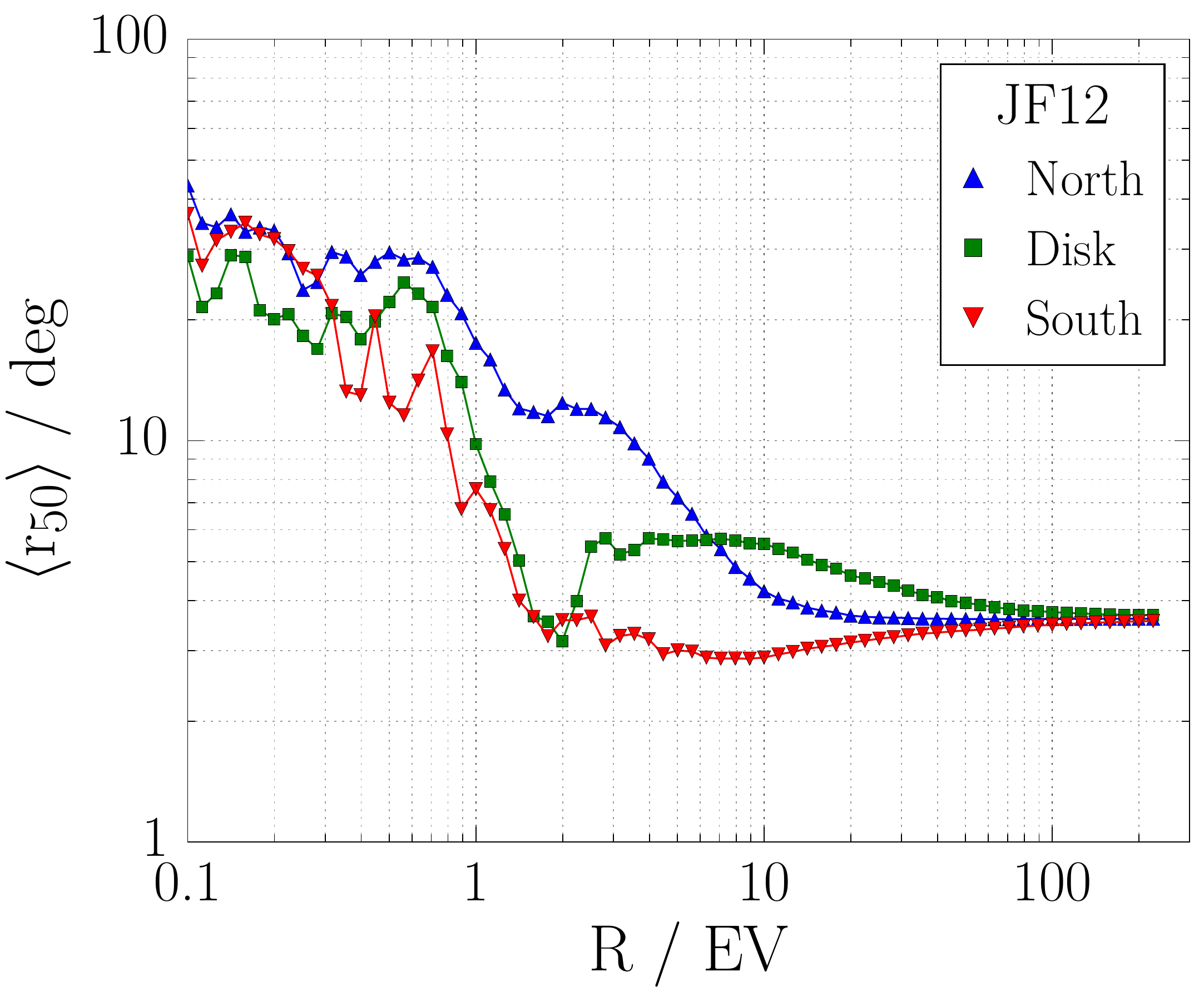}{a)}
\includegraphics[width=0.45\textwidth]{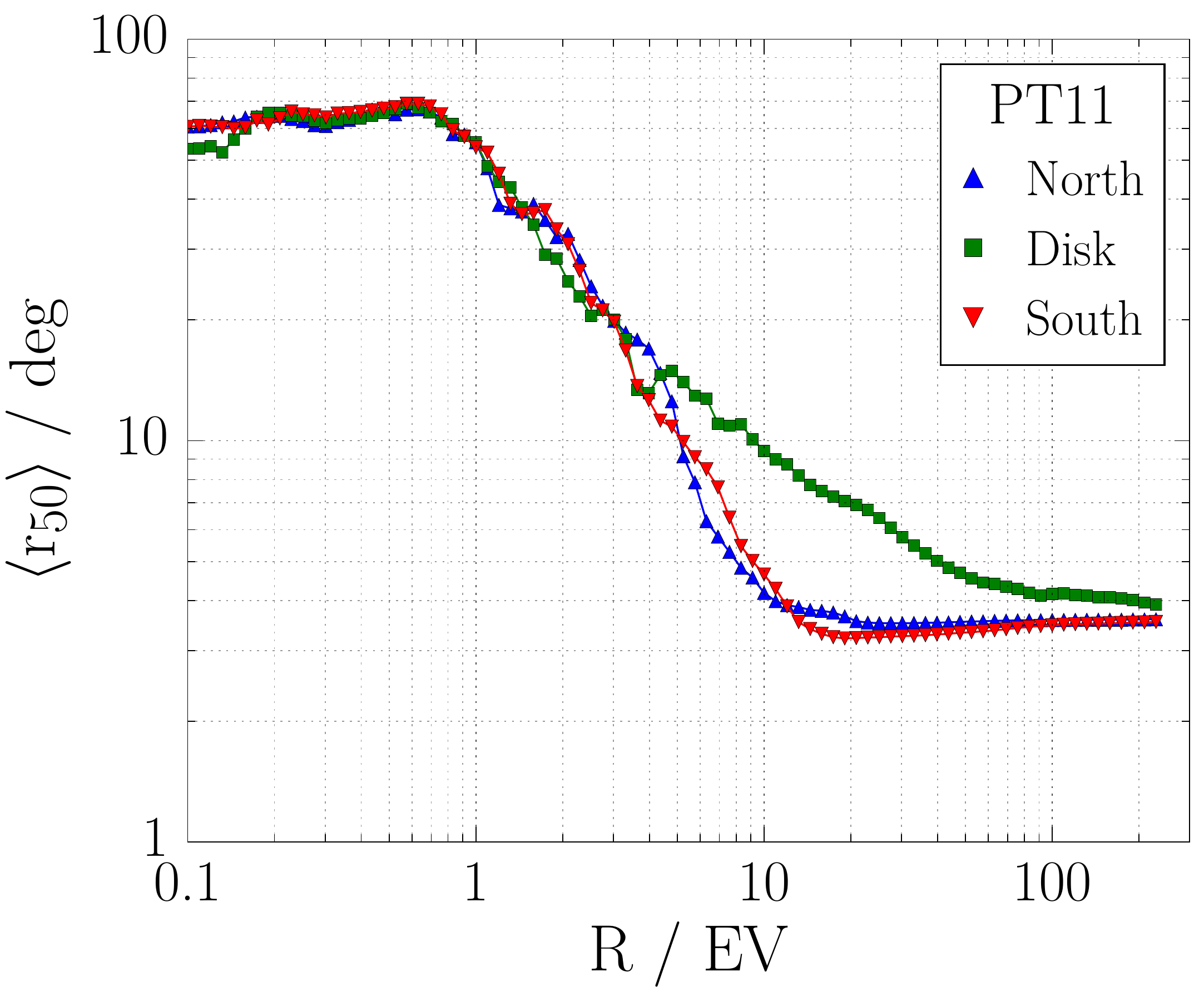}{b)}
\caption{Average extent $\langle r_{50} \rangle$ of arrival probability 
distributions on Earth resulting from cosmic rays incoming to our galaxy with a
fixed Gaussian width of $3$~deg in three regions separated by galactic latitudes 
$\pm 19.5$~deg as a function of rigidity $R$, a) JF12, b) PT11. 
For the exact definition of $r_{50}$ refer to the text.
}
\label{fig:extent}
\end{figure}

In Fig.\ref{fig:extent} we show the extent of the arrival probability distributions 
in terms of the smallest average radius $\langle r_{50} \rangle$ as a function of cosmic
ray rigidity $R$.
Again we show separately the three galactic regions defined above.
As we start with extragalactic directions we use the term ``northern region'' 
to refer to initial directions with galactic latitudes above $19.5$~deg, etc.

The projection of the incoming distribution causes a dispersion 
or a focusing effect, depending on the original cosmic ray direction and its rigidity $R$.
In the southern region (downward-pointing triangles) a focusing effect is visible for 
rigidities around $R\sim 15$~EV.
In contrast, the disk region (square symbols) exhibits dispersion effects up to the largest 
rigidities.
 
Below $R=1$~EV the average extent $\langle r_{50} \rangle$ is large for both fields, 
such that it appears difficult to identify arrival directions on Earth 
from a given extragalactic direction.
The JF12 field exhibits several pronounced features which appear to be specific to the 
JF12 field parametrization. 
Such effects are not visible in the PT11 parametrization where $\langle r_{50} \rangle$ 
appears to increase continuously with reduced rigidity up to $70$~deg.

Above $R=6$~EV both parametrizations lead to similar results with 
a dispersion effect in the disk region (square symbols), and 
dispersion and focusing effects in the southern region (downward-pointing triangles).
The largest extent of the probability region is found for the PT11 field 
corresponding to a dispersion of the initial probability distribution by a factor of $3$
at $R=6$~EV, which may still be acceptable for an analysis of cosmic ray arrival directions.

\subsection{Multiple images}

Extending the above study of dispersion we investigate multiple images arising
in cosmic ray arrival distributions on Earth.
Again cosmic rays from an incoming direction are Fisher-distributed 
with a Gaussian width of $3$~deg. 
For some incoming directions, the small angular deviations within this distribution 
are sufficient to change the arrival direction on Earth substantially, leading to distinct 
maxima in the arrival distributions.
Examples for such multiple images are denoted in Fig.~\ref{fig:images} by the 
red symbols.

We count the number of arrival images by searching for connected areas of arbitrary 
shape in-between which the probability falls below a pre-defined threshold.
For this we coarsen the HEALPix resolution in the arrival distribution from 
originally $1$~deg to $4$~deg and require each pixel to carry at least $20\%$
of the pixel with the maximum arrival probability.
The image multiplicity then arises from counting connected areas that are separated
from one another by at least one pixel below the pre-defined threshold.

\begin{figure}[h]
\includegraphics[width=0.44\textwidth]{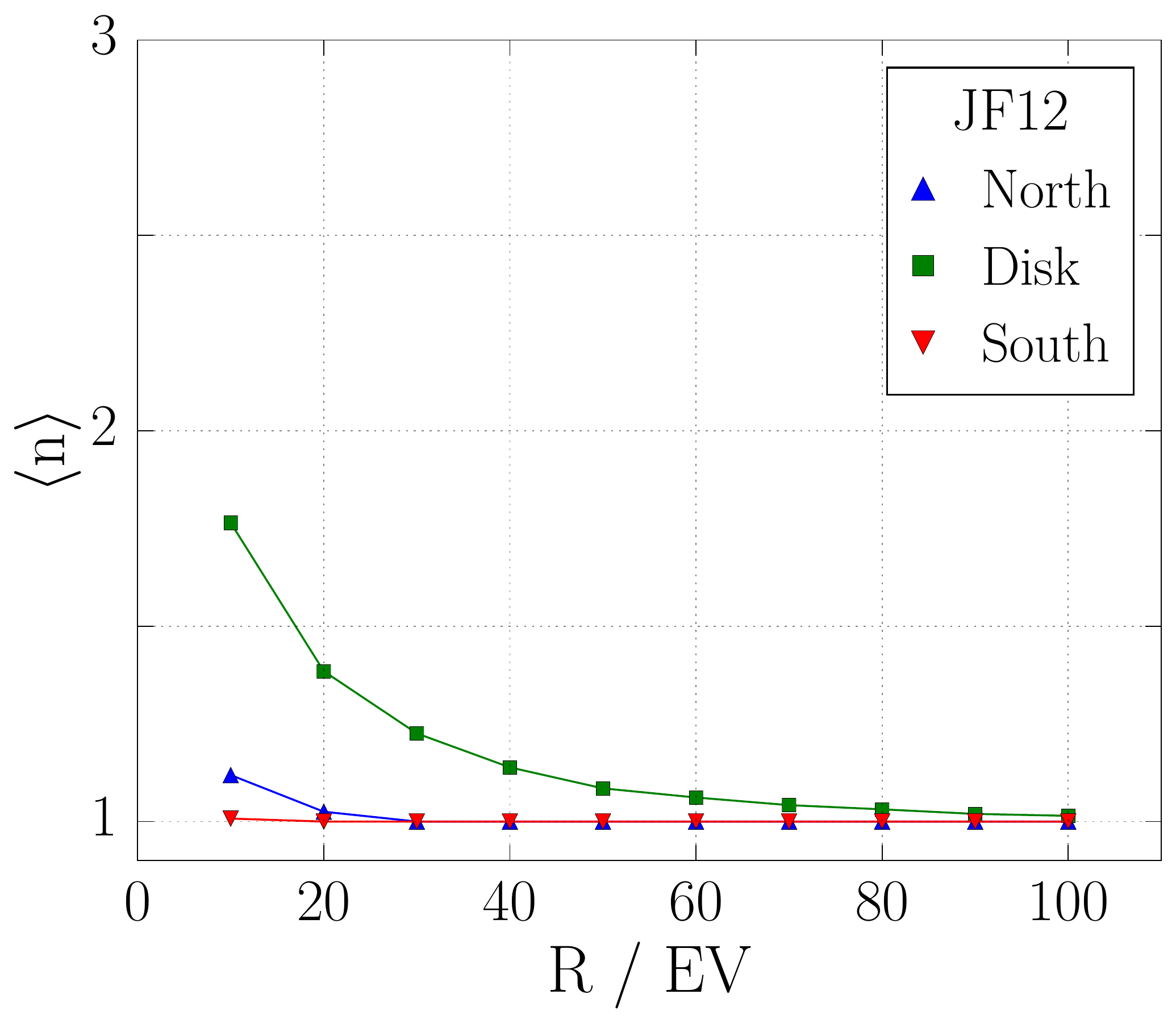}{a)}
\includegraphics[width=0.44\textwidth]{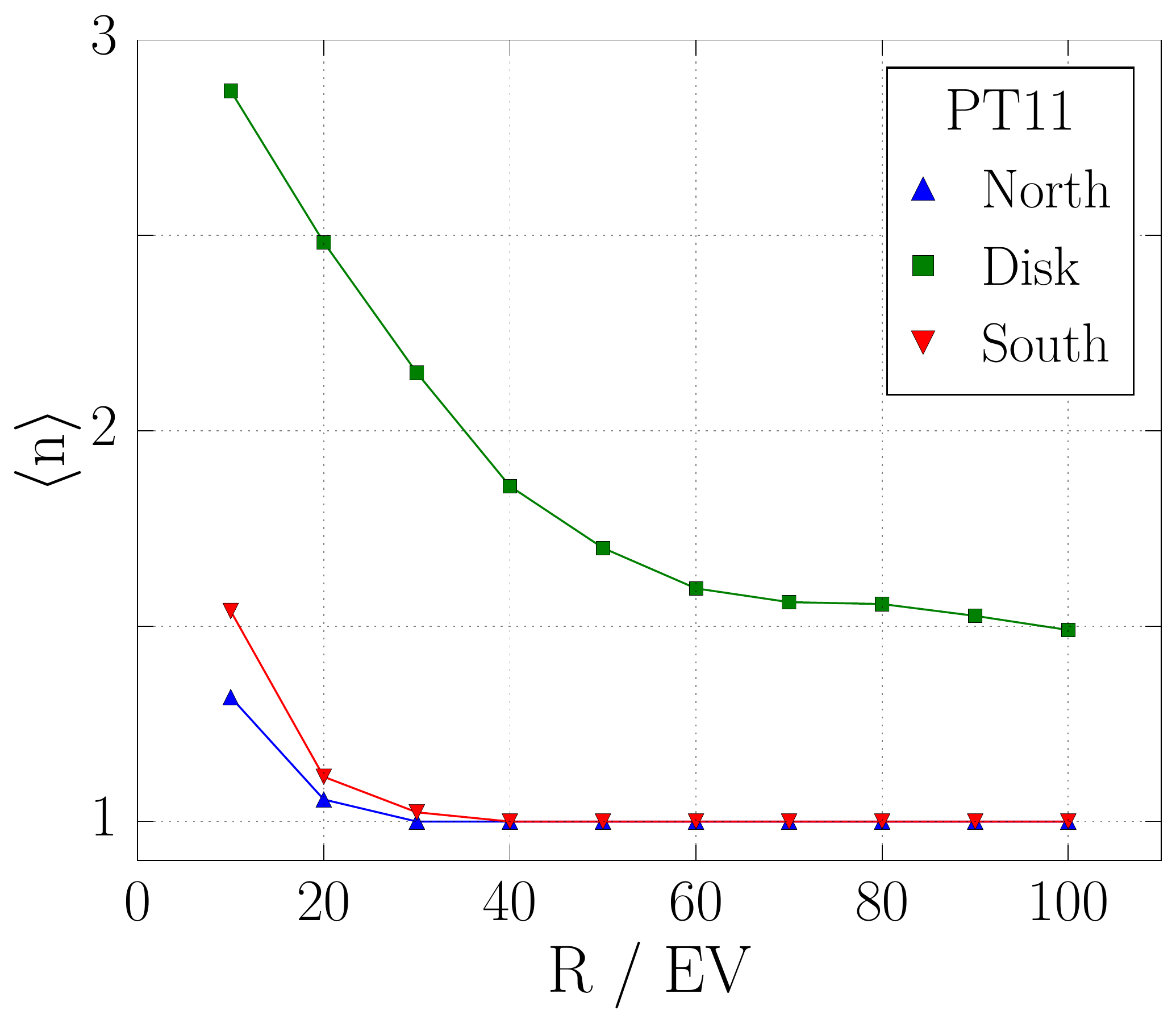}{b)}
\caption{Mean multiplicity $\langle n \rangle$ of images in the arrival probability 
distributions on Earth resulting from cosmic rays incoming to our galaxy with a 
Gaussian width of $3$~deg in three regions separated by galactic latitudes $\pm 19.5$~deg 
as a function of rigidity $R$, a) JF12, b) PT11. 
For the image definition refer to the text.
}
\label{fig:jets}
\end{figure}

In Fig.~\ref{fig:jets} we show the average multiplicity $\langle n \rangle$ 
of arrival images as a function
of cosmic ray rigidity $R$ in the three regions of the galactic sphere separated by
galactic latitudes $\pm 19.5$~deg.
Overall, the image multiplicity decreases with increasing rigidity.
The multiplicity arising from the PT11 parametrization (Fig.~\ref{fig:jets}b)
appears to exceed that of the JF12 parameterization (Fig.~\ref{fig:jets}a).

For directions in the northern and southern regions, typically $1$ image 
of the arrival direction arises. 
In the disk region, however, multiple images appear even for cosmic rays with large rigidity.
Here the image multiplicity is especially large for the PT11 parameterization
(Fig.~\ref{fig:jets}b, square symbols) which is related to the pronounced halo field
visualized in Fig. \ref{fig:parameterization}b.

Multiple images reduce the predictive power of cosmic ray arrival directions
and may require additional selection depending on the individual analysis.
As criteria to reduce image multiplicity both cosmic ray rigidity and a selection 
of incoming directions away from the galactic disk region are relevant.

\subsection{Field transparency}

Related to the above study on dispersion we investigate extragalactic directions
causing a relatively enhanced flux of cosmic rays on Earth \citep{Harari2000}, and
directions for which the arrival probability disappears as no simulated trajectory leads 
to Earth.
These effects have a direct impact on the visibility of a source by cosmic ray messengers, 
and the luminosity required for observation on Earth. 
Examples of varying transparency of the field depending on the incoming directions 
are shown in Fig.~\ref{fig:images}.

\begin{figure}[h]
\includegraphics[width=0.45\textwidth]{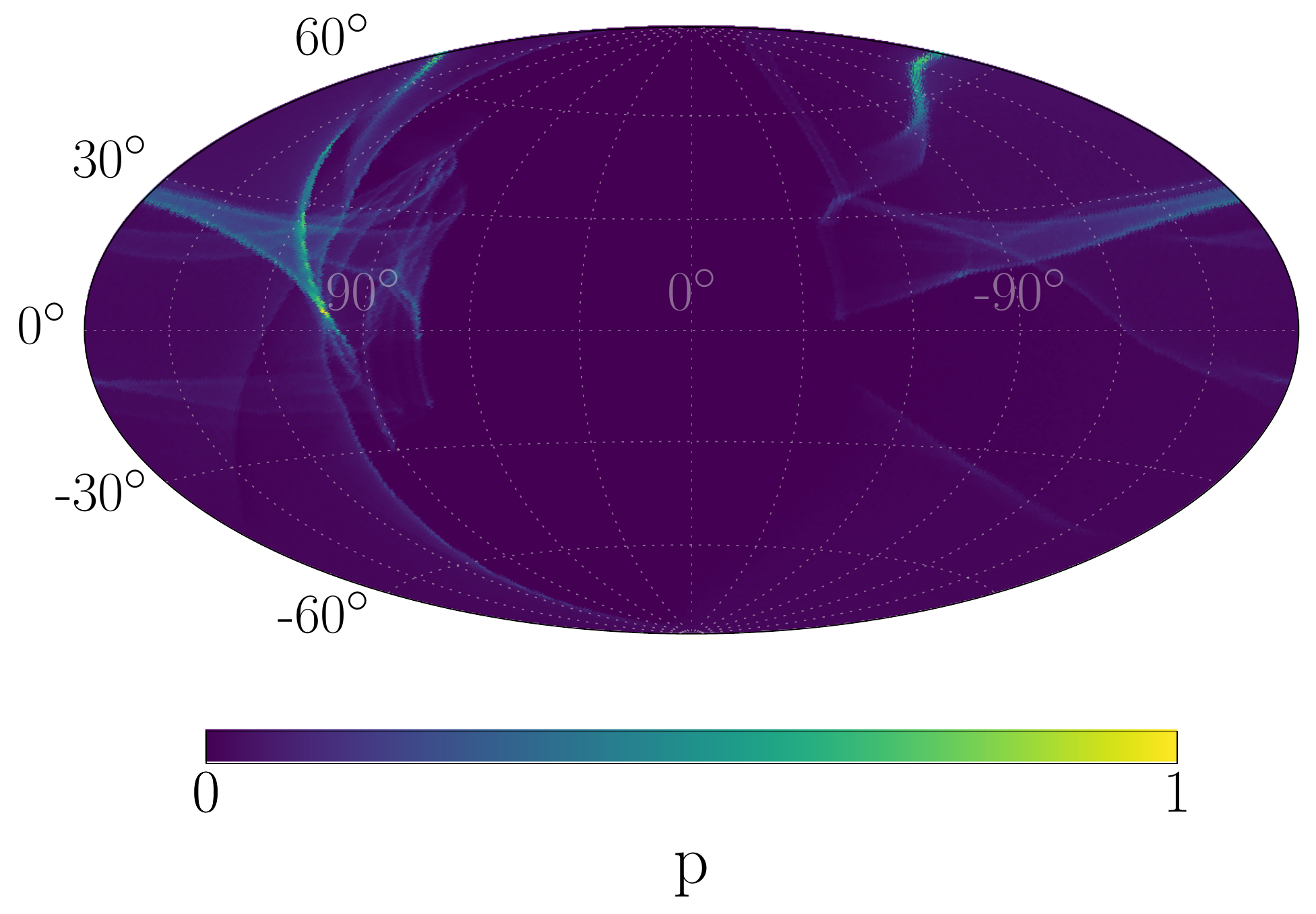}{a)}
\includegraphics[width=0.45\textwidth]{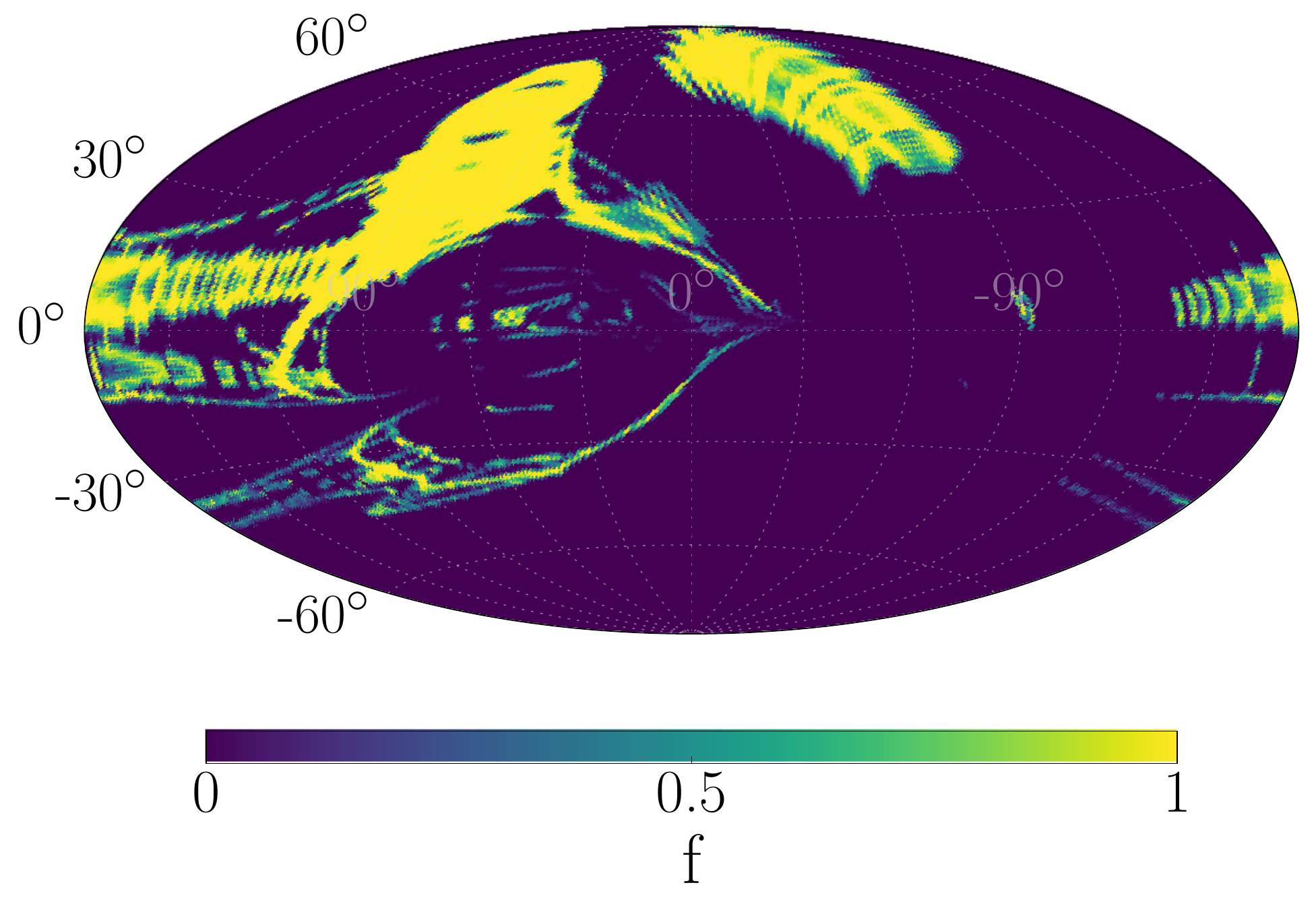}{b)}
\caption{a) Probability $p$ of observing a cosmic ray with rigidity $R=6$~EV on Earth 
as a function of the incoming direction to our galaxy (JF12).
b) Flux $f$ on Earth originating from the $1\%$-percentile of the
directions with the largest arrival probabilities in Fig. a).
}
\label{fig:focusing}
\end{figure}

To demonstrate the enhanced flux of a few extragalactic directions we show 
in Fig.~\ref{fig:focusing}a for incoming cosmic rays with rigidity $R=6$~EV
the probability $p$ of arriving on Earth coded in color.
The lightly colored regions indicate incoming cosmic ray directions with a high
probability of observation on Earth.

In order to quantify flux enhancement we organize the incoming directions 
(binned in $N_{pix}=49,152$ pixels of $1$~deg) according to their arrival 
probabilities $p_j$ on Earth and select the leading $k$ directions.
These incoming directions cover a solid angular region of 
$\Omega=(k/N_{pix})\cdot 4\pi$ and provide a relative flux contribution of
$F(k)=\sum_{j=1}^k p_j/\sum_{j=1}^{N_{pix}} p_j$.

In Fig.~\ref{fig:focusing}b we show for the above example the observed flux $f$ 
of cosmic rays originating from the $1\%$ incoming directions with the 
highest probabilities indicated in Fig.~\ref{fig:focusing}a, when assuming 
an isotropic extragalactic flux.
We find that these few incoming directions cause a wide spread distribution and 
contribute $F=30\%$ to the observed flux.

\begin{figure}[h]
\includegraphics[width=0.45\textwidth]{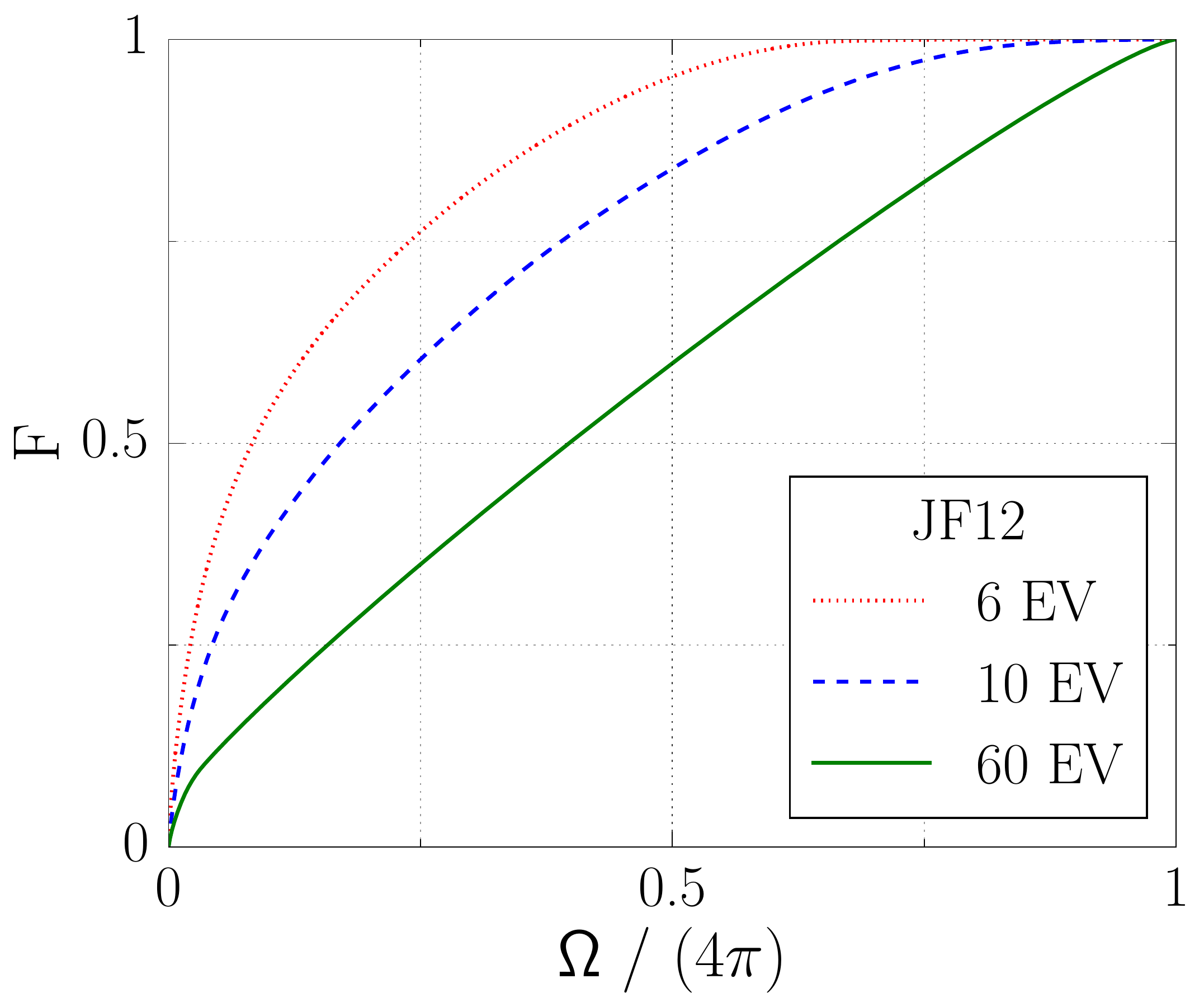}{a)}
\includegraphics[width=0.45\textwidth]{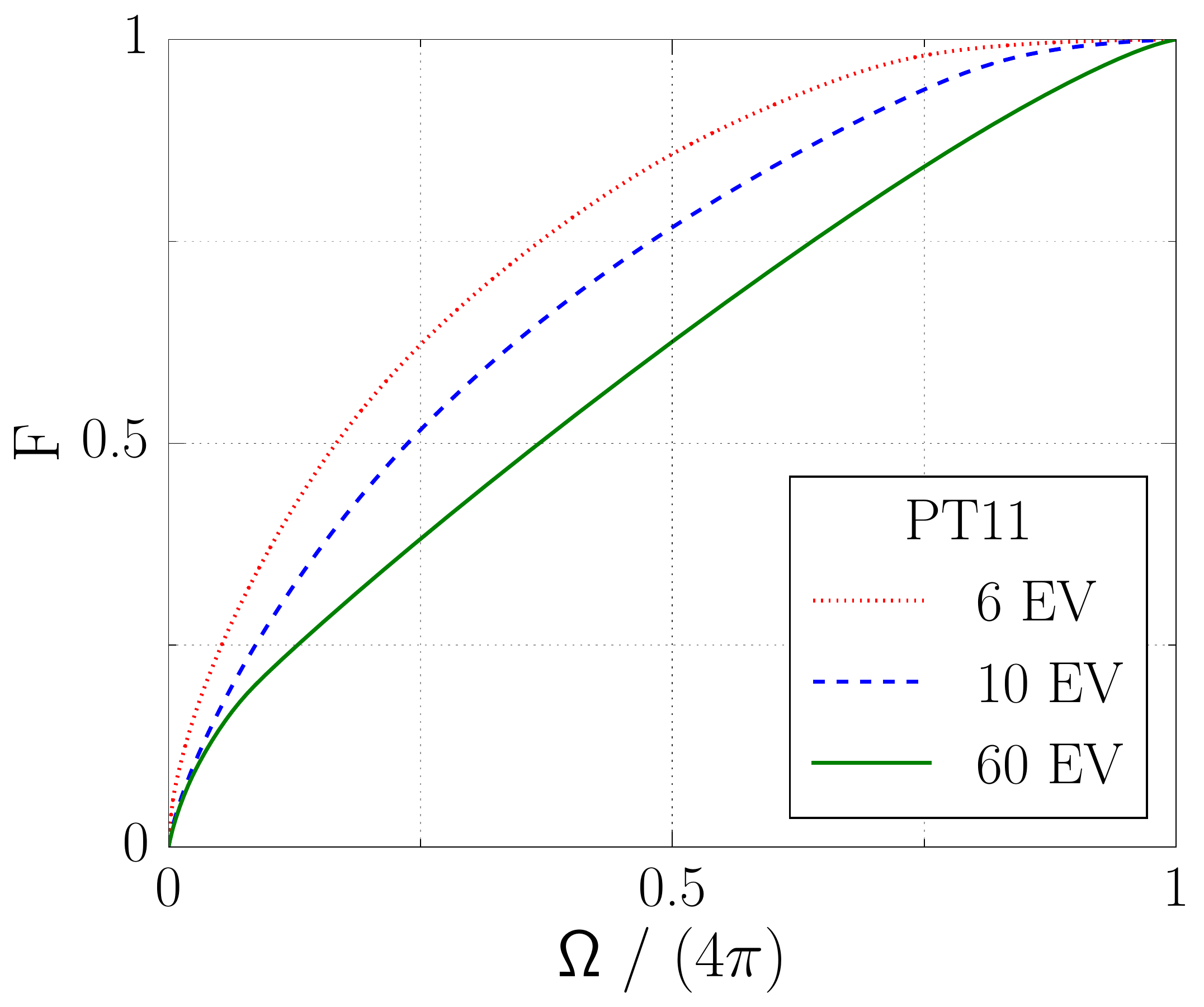}{b)}
\caption{Integrated relative flux $F$ of observed cosmic rays originating from the
incoming directions with the highest arrival probabilities
covering $\Omega/(4\pi)$ of the sky, a) JF12, b) PT11.
}
\label{fig:int_flux}
\end{figure}

In a more general approach we show in Fig.~\ref{fig:int_flux} the relative flux $F$
of observed cosmic rays as a function of the solid angular region $\Omega/(4\pi)$ 
covered by the incoming directions with the highest arrival probabilities.
At low rigidity $R=6$~EV in the JF12 parameterization
$95\%$ of the cosmic ray flux on Earth is caused by about $50\%$ 
of the extragalactic directions  (Fig.~\ref{fig:int_flux}a).
With increasing rigidity all extragalactic directions contribute equally to the flux
on Earth.
The PT11 parameterization yields similar results as shown in Fig.~\ref{fig:int_flux}b,
however, with less inhomogeneity at low rigidity $R$.

\begin{figure}[h]
\includegraphics[width=0.45\textwidth]{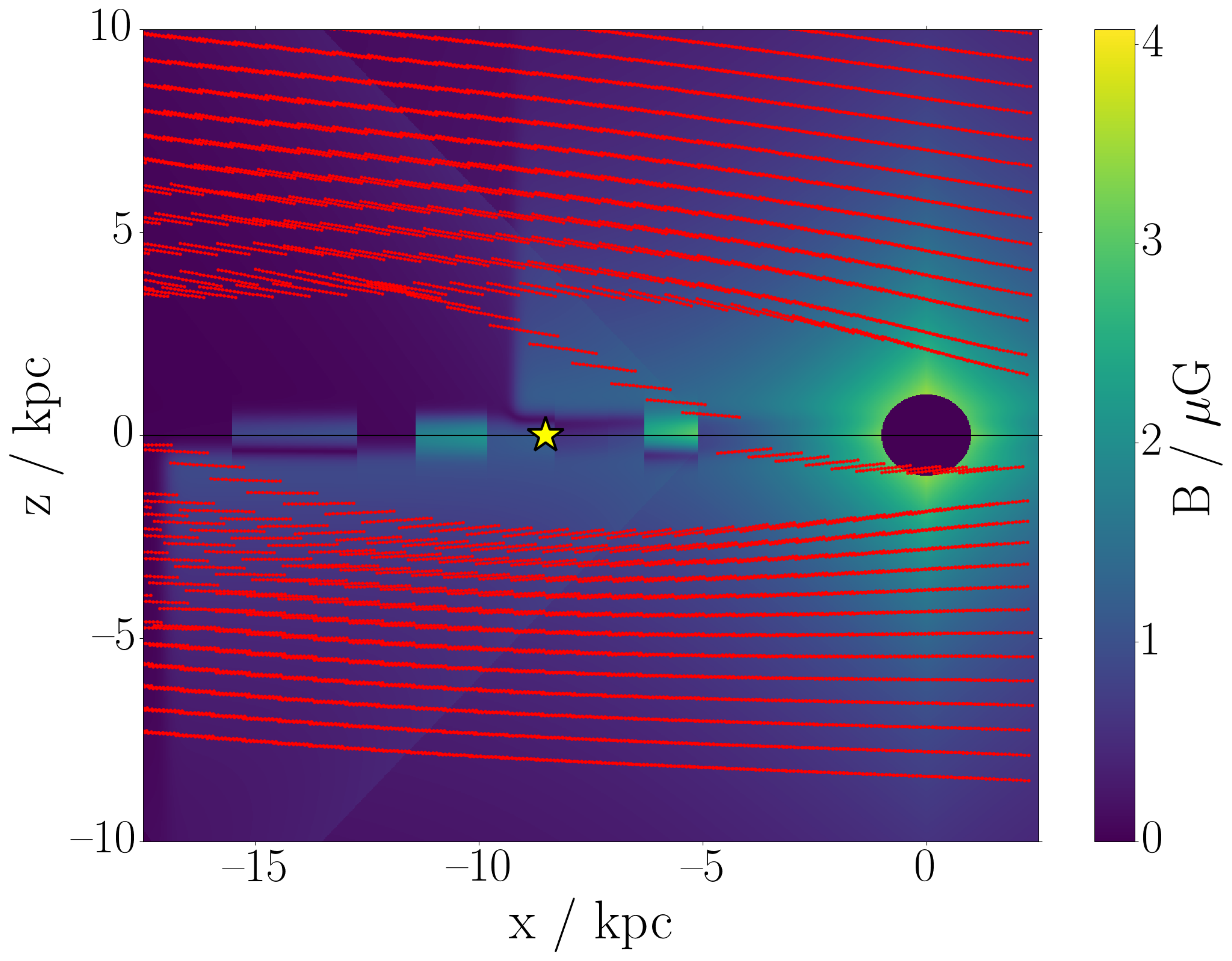}{a)}
\includegraphics[width=0.44\textwidth]{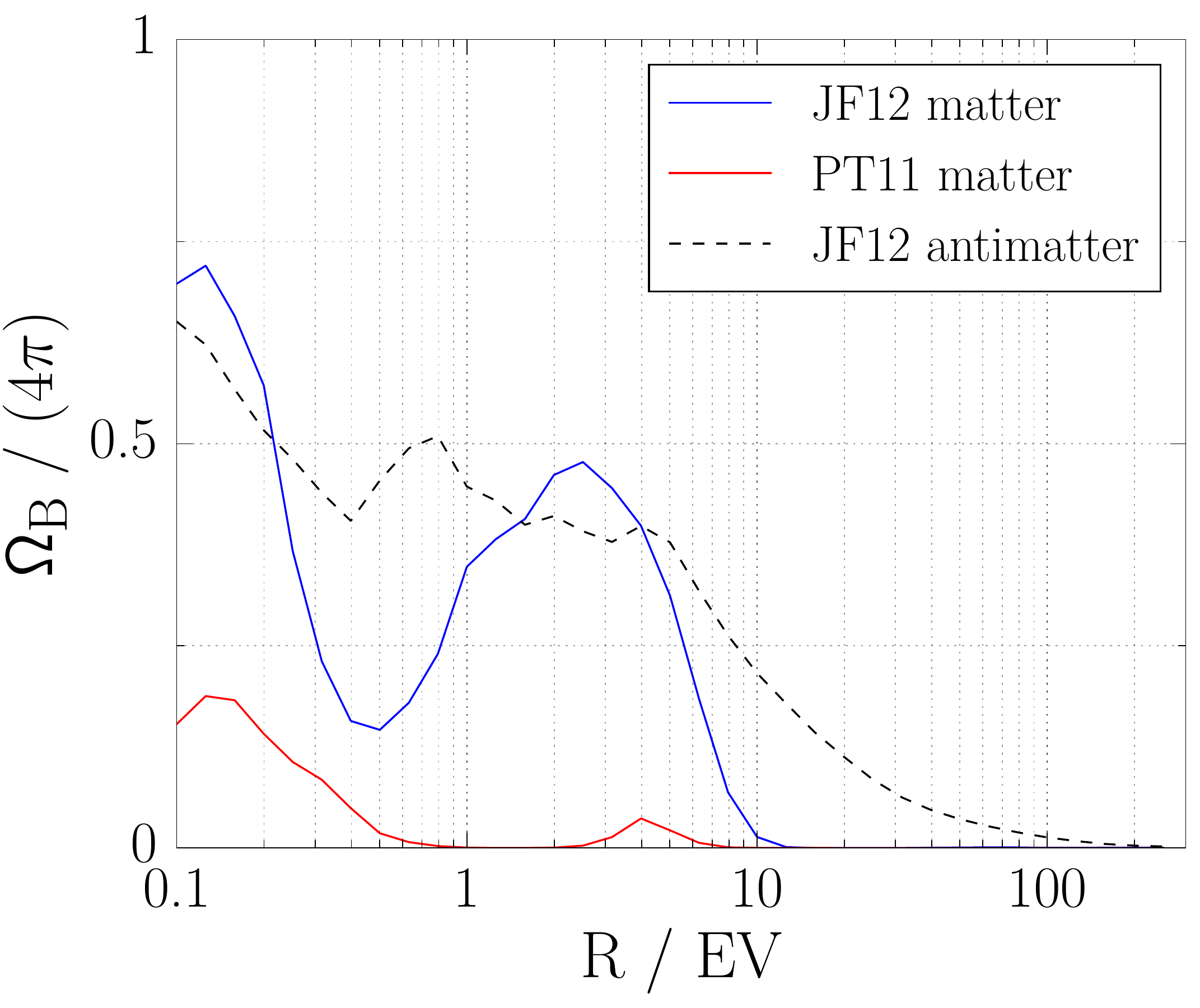}{b)}
\caption{a) Example trajectories of cosmic antimatter with rigidity $R=60$~EV 
incoming to the JF12 field from the same direction that 
miss the solar system (yellow star). The trajectories are shown in a thin slice 
of $\pm 500$~pc around the $x$-$z$-plane. 
b) Directions with unobserved cosmic matter (full curves) and antimatter 
(dashed curve) in terms of covered solid angle $\Omega_B$ as a function 
of rigidity $R$ (blue curve JF12, red curve PT11).
}
\label{fig:antimatter}
\end{figure}

In contrast, the flux from certain extragalactic directions is not only  
suppressed but can even disappear, as no path leads to observation on Earth.
As an example we show in Fig.~\ref{fig:antimatter}a example trajectories of 
cosmic antimatter with rigidity $R=60$~EV from the same extragalactic direction
traversing a thin slice of $\pm 500$~pc around the $x$-$z$-plane in the galactic coordinate system.
The trajectories are expected to miss the solar environment (marked by the yellow star).
To enable this demonstration for antimatter we produced a separate set of lenses by backtracking matter 
particles in the JF12 field.

In general, both matter and antimatter particles entering the galaxy exhibit directions 
with a negligible arrival probability on Earth.
In Fig.~\ref{fig:antimatter}b we show the fraction of the sky in terms of solid angles
$\Omega_B$, where none of $100$ simulated cosmic rays reached Earth as a function of their rigidity $R$. 
For antimatter trajectories in the JF12 field, invisible directions appear 
at small and large rigidities (dashed curve).

For matter particles of rigidity $R=6$~EV traversing the JF12 field, the invisible sky fraction is 
about $20\%$ and disappears above $R\sim 10$~EV (blue curve).
In contrast, for the PT11 field the invisible fraction of the sky for matter particles 
appears to be generally small (red curve).

At cosmic ray rigidities of $R=6$~EV, variations in the field transparency from specific 
extragalactic directions lead to relative suppression and enhancement 
effects which may eventually require corrections in individual arrival direction analyses.
The variations are strongly reduced with increasing cosmic ray rigidity.

\subsection{Small-scale random field \label{sec:turbulent} }

Beyond the regular large-scale component, recent galactic magnetic field models also 
contain small-scale random structures which are motivated e.g. by supernovae
\citep{Pshirkov2013, Jansson2012b, Beck2014}.
Such local disturbances are expected to cause a randomly oriented field component
which introduces uncertainties in the predicted arrival directions of extragalactic 
cosmic rays.
We expect this impact to be small, since the direction of the random component 
changes on a scale that is substantially smaller than the gyroradius of 
cosmic rays constrained in the galaxy. 

To investigate the influence of such random fields we compare two different realizations of 
the so-called striated and turbulent random components as described in \citep{Jansson2012b} 
with a coherence length of $\lambda = 60$~pc. 
Cosmic rays are then deflected in both the regular JF12 and the first random field realization,
and in the regular JF12 and the second random field, respectively.

As we aim to investigate the influence of the random fields on arrival directions 
we use the lensing technique.
To obtain the most probable arrival direction on Earth we use the same techniques described
above when studying the dispersion of the probability distribution (section \ref{sec:dispersion}).
We calculate the radius $r_{50}$ containing $50\%$ of the arrival probabilities, and 
use the center of the pixel with the smallest radius $r_{50}$ as the expected arrival 
direction.

\begin{figure}
\includegraphics[width=0.4\textwidth]{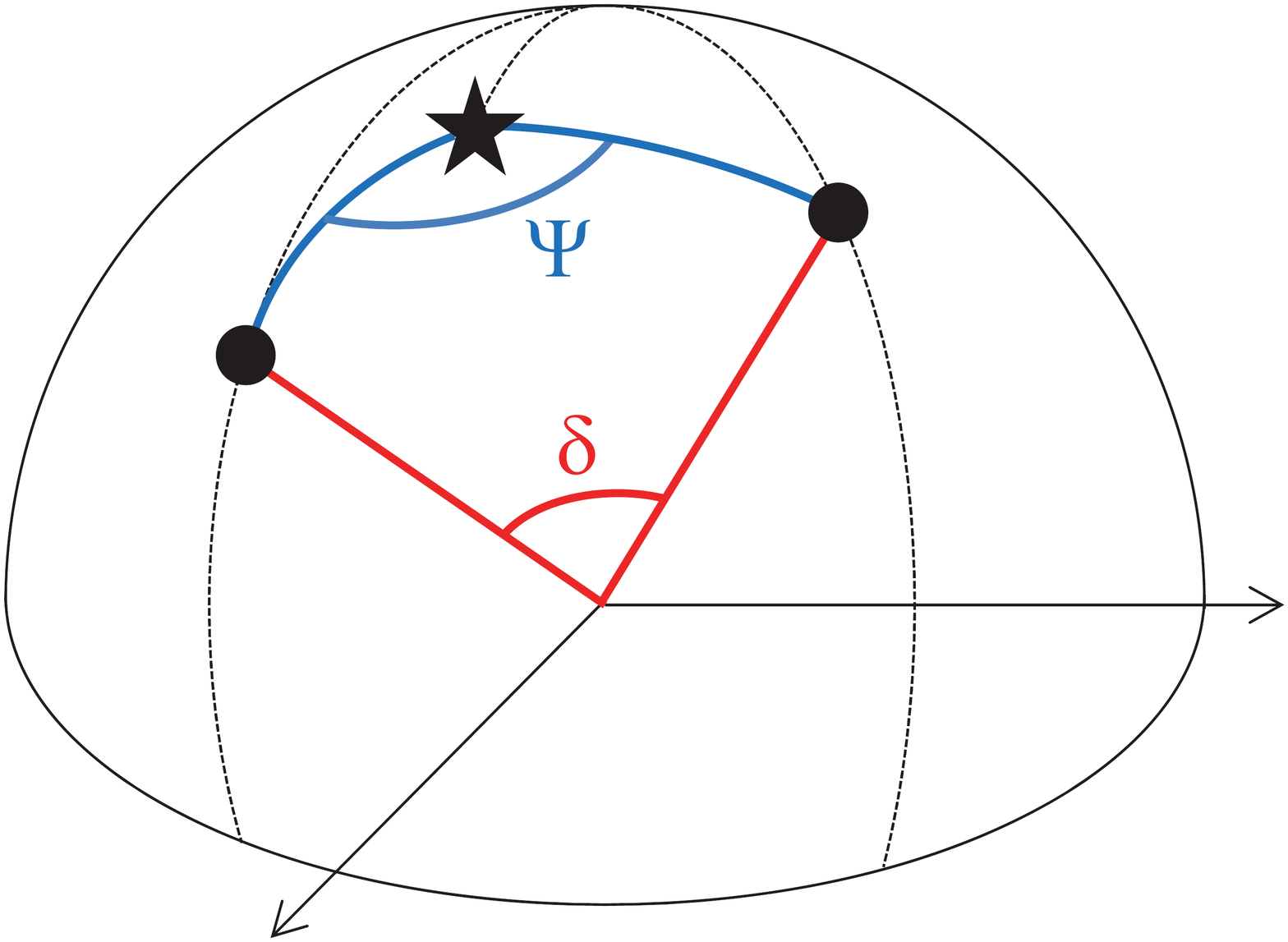}{a)}
\includegraphics[width=0.45\textwidth]{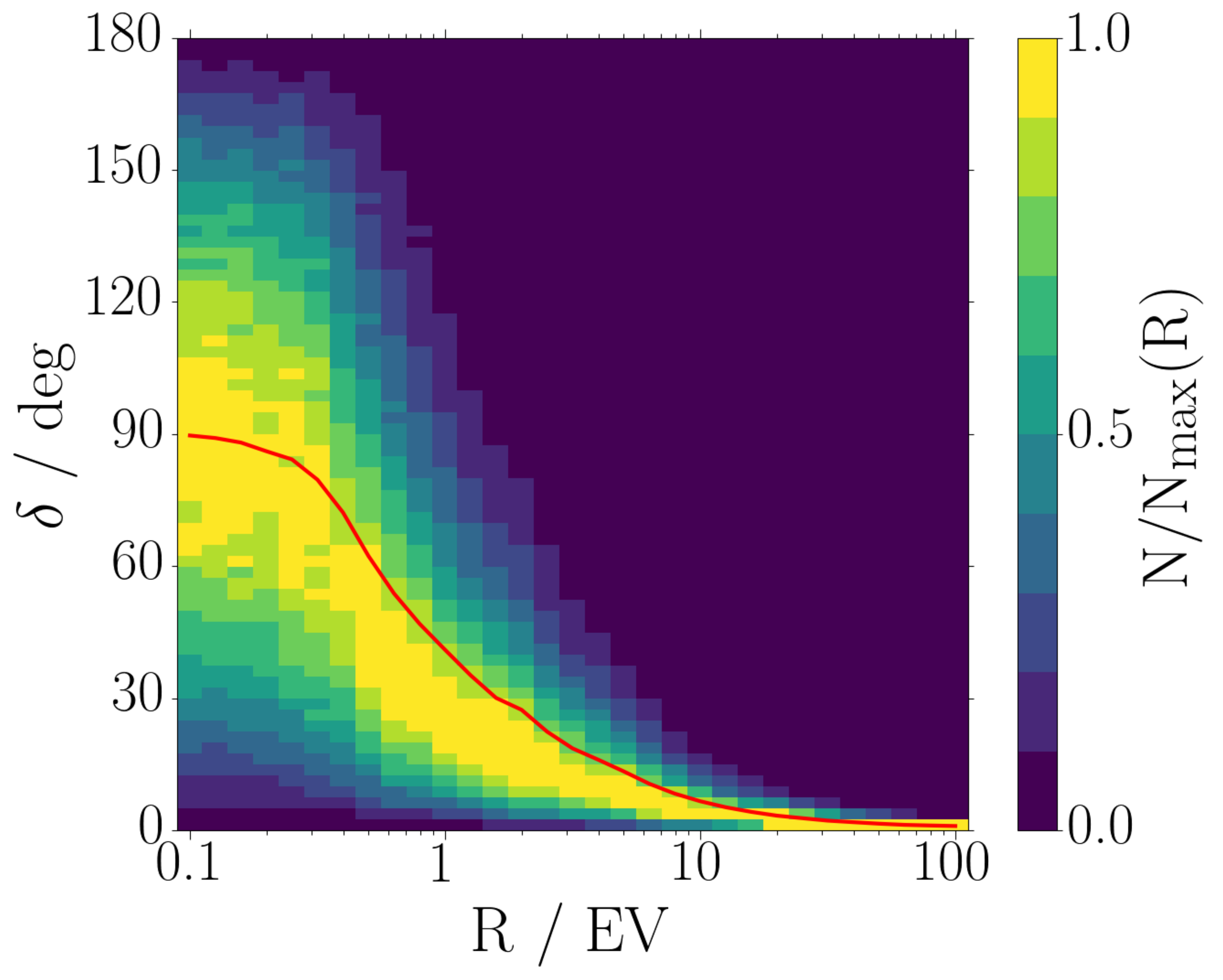}{b)}
\caption{a) Angular distance $\delta$ between two cosmic ray arrival directions (circular symbols) 
from an incoming direction (star symbol), and directional difference $\Psi$ 
of the deflections (tangential to the sphere) resulting from different magnetic field 
orientations.
b) Angular distances $\delta$ between the arrival directions using two different 
realizations of the striated and turbulent random components in addition to the
regular JF12 field parametrization as a function of rigidity $R$. 
The median is depicted by the curve.}
\label{fig:delta}
\end{figure}

In Fig.\ref{fig:delta}a we sketch the angular distance $\delta$ between the arrival 
directions resulting from the two random field realizations.
Here the incoming cosmic ray direction to the galaxy is indicated by the star symbol,
and the two alternative arrival directions are denoted by the circular symbols.

In Fig.\ref{fig:delta}b we show the angular distances $\delta$ between the arrival directions on Earth
using the two realizations of the random fields as a function of cosmic ray rigidity $R$.
The red curve indicates the median values.
For rigidity $R=6$~EV the uncertainty in the arrival directions is below $10$~deg for $50\%$
of the cosmic rays, and rarely extends to more than $90$~deg.

As expected, the impact of the random field is significantly smaller compared to the 
deflections of the cosmic rays arising from the regular galactic field (Fig.\ref{fig:angular}).
However, the uncertainties generated by such random deflections are sufficiently sizable 
to consider optimization of the rigidity threshold for individual arrival 
direction analyses.

\subsection{Summary of the field impact on cosmic ray arrival}

In order to work in a phase space region in which cosmic ray deflections can be controlled
at least in a probabilistic way a minimum cosmic ray rigidity of $R=6$~EV is recommended.
This value results from avoiding deflections leading to a bend of $90$~deg which overlaps
with cosmic ray diffusion.
All other distributions presented above on the dispersion of arrival probability 
distributions and multiple images, variance in field transparency, and uncertainties 
due to random field components are in accordance with this minimal rigidity value.

For a typical large-scale analysis the $R=6$~EV rigidity threshold may be sufficient.
However, depending on individual analysis requirements, the above key distributions
may help to determine whether the rigidity threshold needs to be adjusted to larger values,
or whether restrictions to incoming directions aside the galactic disk region need to be 
introduced.

\section{Influence of field uncertainties on cosmic ray arrival}

For a typical point source search, the reliability of galactic field corrections are
of utmost importance.
As a first step we compare directly the cosmic ray deflections resulting from the 
two field parameterizations PT11 and JF12.

In order to exemplify the impact of these differences on point source searches 
we simulate a typical corresponding analysis.
We determine the discovery potential when the true galactic field is known, and
quantify the reduced discovery potential when taking into account the different deflections
of the two field parameterizations.

\subsection{Comparison of deflection angles}

Uncertainties in the current knowledge of the galactic field can be obtained 
to some extent from the different arrival distributions of the two field 
parameterizations PT11 and JF12.
Note that their fields are not completely independent regarding the overlap in the 
measurements constraining their fits, and usage of similar electron density 
distributions.
However, the two parameterizations follow different ansatzes and include 
disjoint measurements, such that a direct comparison of cosmic ray deflections 
at least gives an idea of limited knowledge in the magnetic field.

For cosmic rays originating from the identical extragalactic direction $(l ,b)$
we investigate the different deflections resulting from the two field parameterizations.
We use the above-mentioned lensing technique to deliver arrival probability 
distributions.
As outlined in section \ref{sec:turbulent},
we calculate the most probable arrival direction on Earth 
by using the center of the pixel with the smallest radius $r_{50}$ containing $50\%$ of the 
arrival probabilities.
This gives the expected arrival directions $(l^\prime, b^\prime)$ 
for the JF12 parametrization and for the PT11 parametrization, respectively.

We first study the different directions of cosmic ray deflections 
by their azimuthal angular distance $\Psi$ which is measured tangentially 
to the sphere (see Fig.\ref{fig:delta}a).
These directional differences reflect different field orientations 
of the two parameterizations.

\begin{figure}[h]
\includegraphics[width=0.44\textwidth]{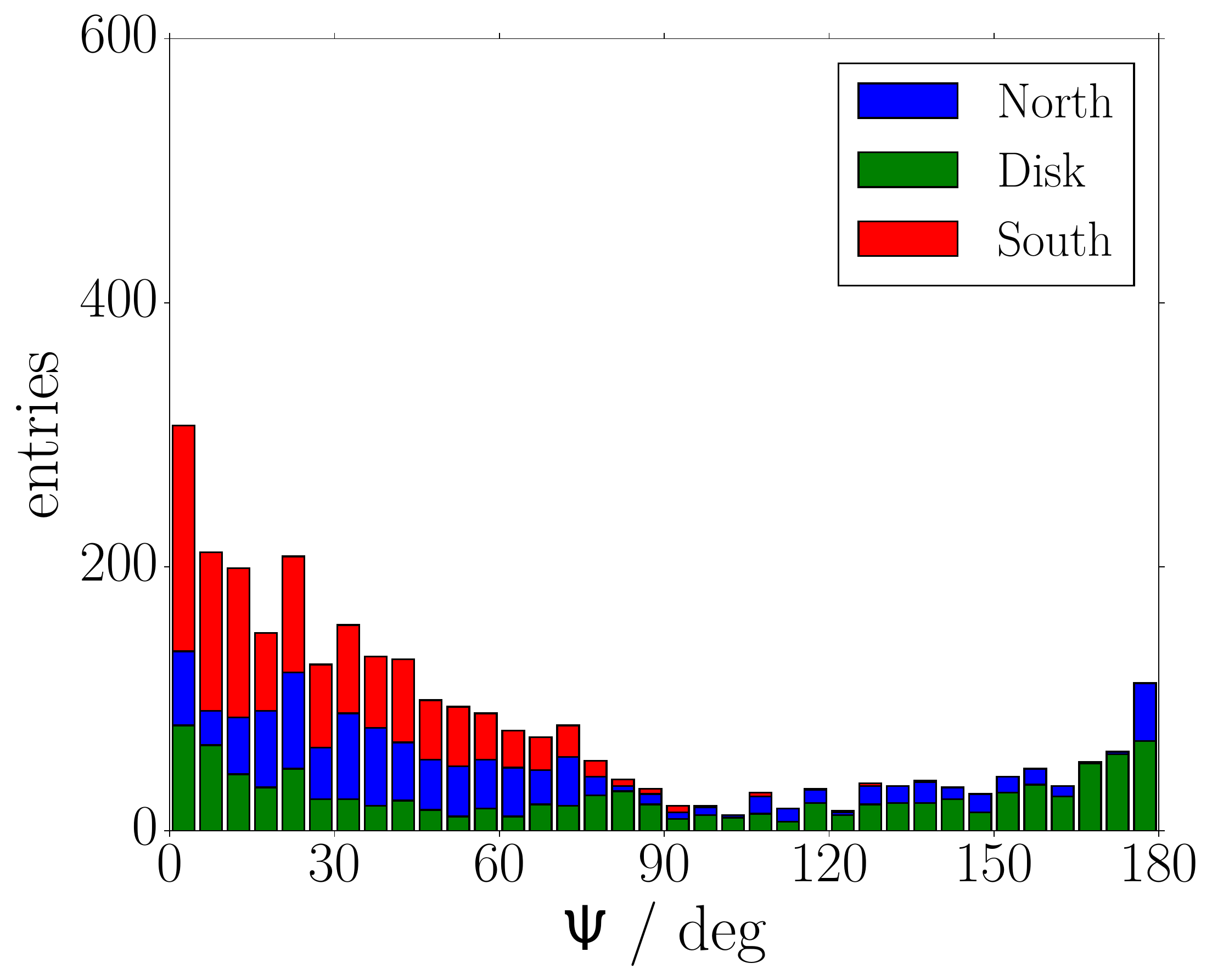}{a)}
\includegraphics[width=0.45\textwidth]{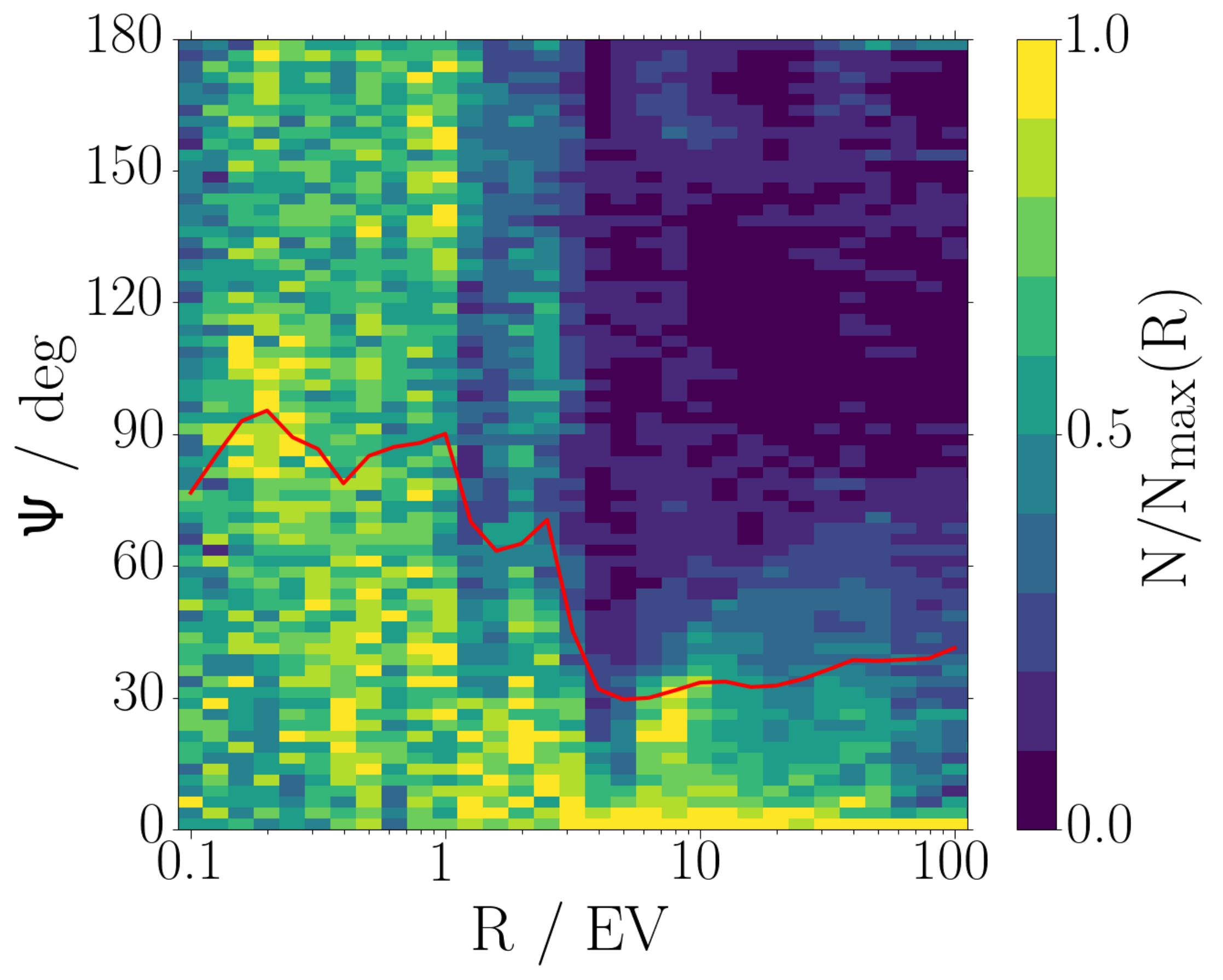}{b)}
\caption{Difference $\Psi$ in the directions of the deflections between JF12 and PT11, 
a) for rigidity $R=60$~EV in the three regions separated by galactic latitudes $\pm 19.5$~deg,
b) as a function of rigidity.
The curve depicts the median values.
}
\label{fig:psi}
\end{figure}

In Fig.\ref{fig:psi}a we show the directional difference $\Psi$ for cosmic ray
rigidity $R=60$~EV for the three regions separated by galactic latitudes $\pm 19.5$~deg.
For the southern region almost all directions of the deflections are within $90$~deg (red histogram).
In the northern region most of the incoming directions show $\Psi<90$~deg (blue histogram).
About $1/5$ of the incoming directions exhibit $\Psi>90$~deg, i.e. here the directions are 
nearly opposite.

In the disk region the differences between the two fields are large as half of the incoming 
directions are within $90$~deg, and the other half has directional differences above $90$~deg 
(green histogram).

\begin{figure}[h]
\includegraphics[width=0.44\textwidth]{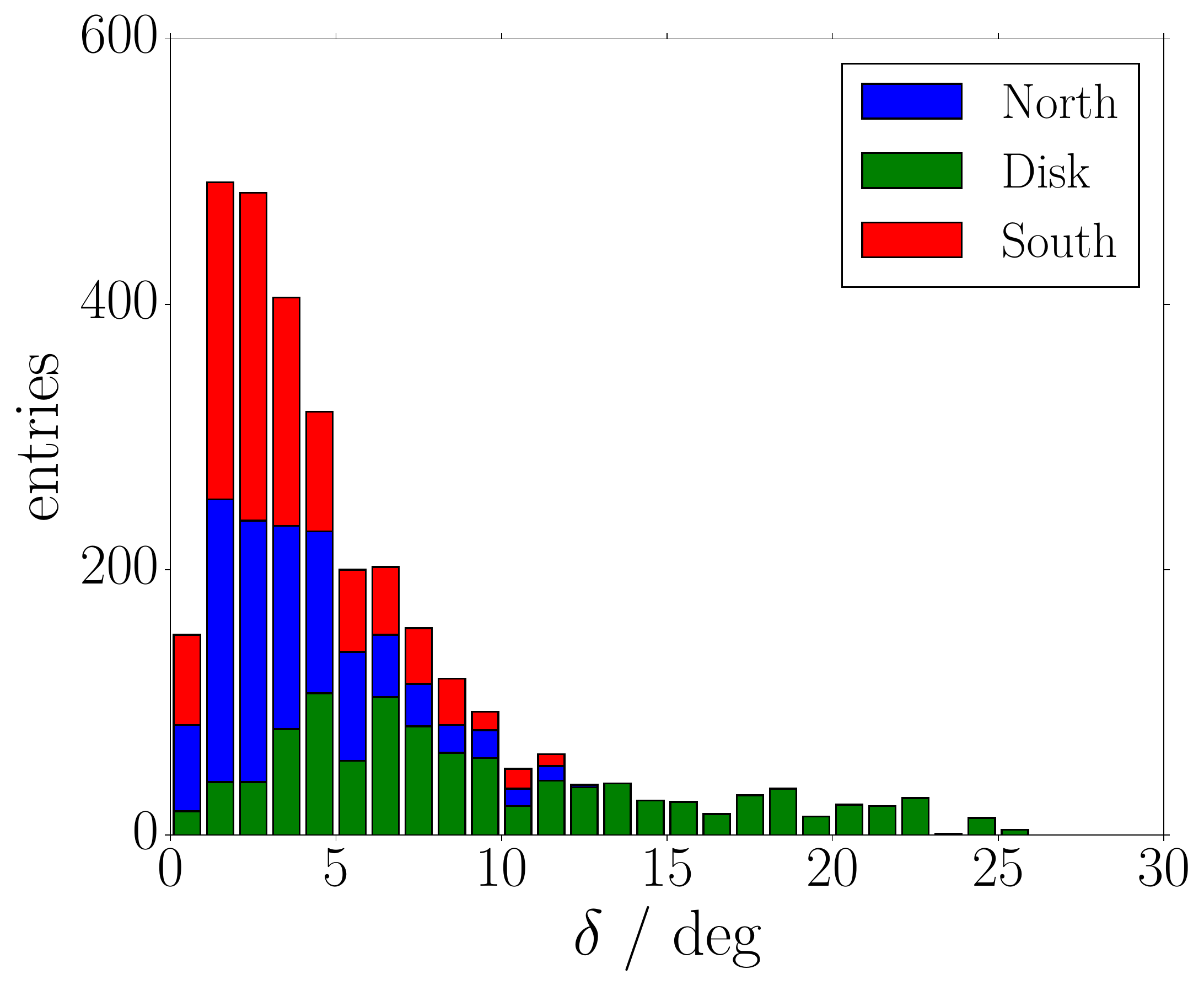}{a)}
\includegraphics[width=0.45\textwidth]{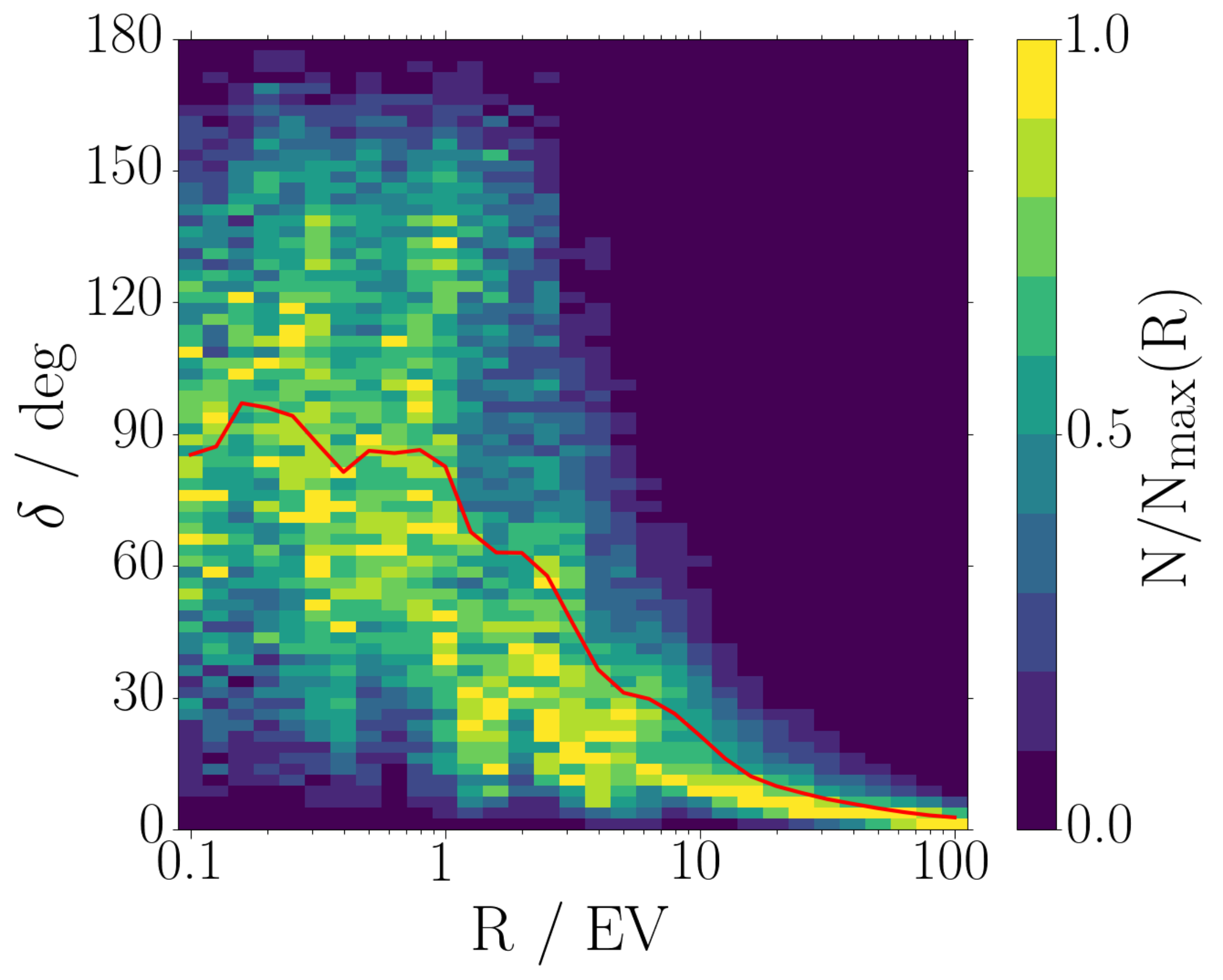}{b)}
\caption{Angular distance $\delta$ of cosmic rays after deflections in JF12 and PT11, 
a) for rigidity $R=60$~EV in the three regions separated by galactic latitudes $\pm 19.5$~deg,
b) as a function of rigidity.
The curve depicts the median values.
}
\label{fig:comparison}
\end{figure}

In Fig.\ref{fig:psi}b we show the rigidity dependence of the directional difference $\Psi$
for all regions where the curve represents the median values.
For all rigidities above $R=6$~EV, the most likely differences in the directions of the 
deflections are below $10$~deg, and for $50\%$ of the cosmic rays the field directions are within $40$~deg.

The second important aspect of the field differences is the absolute angular distance 
in the arrival directions resulting from the PT11 and JF12 parameterizations which are 
denoted by $\delta$ (see Fig.\ref{fig:delta}a).

In Fig.\ref{fig:comparison}a we show the angular distance $\delta$ for cosmic rays with 
rigidity $R=60$~EV for the three regions separated by galactic latitudes $\pm 19.5$~deg.
For the northern and southern regions the angular distance between the two 
parametrizations is below $\delta = 5$~deg for $3/4$ of the incoming directions
(blue, red histograms).
 
In the disk region, only $1/3$ of the arrival directions show angular distances 
below $\delta = 5$~deg, while the majority of incoming directions have larger 
angular distances up to $\delta = 30$~deg.

In Fig.\ref{fig:comparison}b we show the angular distance $\delta$ for all regions 
as a function of cosmic ray rigidity.
The curve represents the median values.
For cosmic rays with small rigidity $R=6$~EV, half of them result at an angular 
difference below $30$~deg.
However, there is a long tail towards large angular distances resulting from the two field
parameterizations.

Although sizable differences in the directional characteristics and magnitudes of 
the two field parameterizations exist, their influence is sufficiently reduced at 
large cosmic ray rigidity.
For example, at $R=60$~EV, the absolute deflection angles $\beta$ as well as the 
angular distances $\delta$ arising from the two fields become consistently small 
for most incoming directions.
Further studies show that in the northern and southern regions the median ratio of the 
angular distance $\delta$ and the averaged deflection 
$(\beta({\rm JF12})+\beta({\rm PT11}))/2$ remains approximately flat for rigidities 
$R>10$~EV.

\subsection{Comparison of arrival directions}

\begin{figure}[h]
\includegraphics[width=0.44\textwidth]{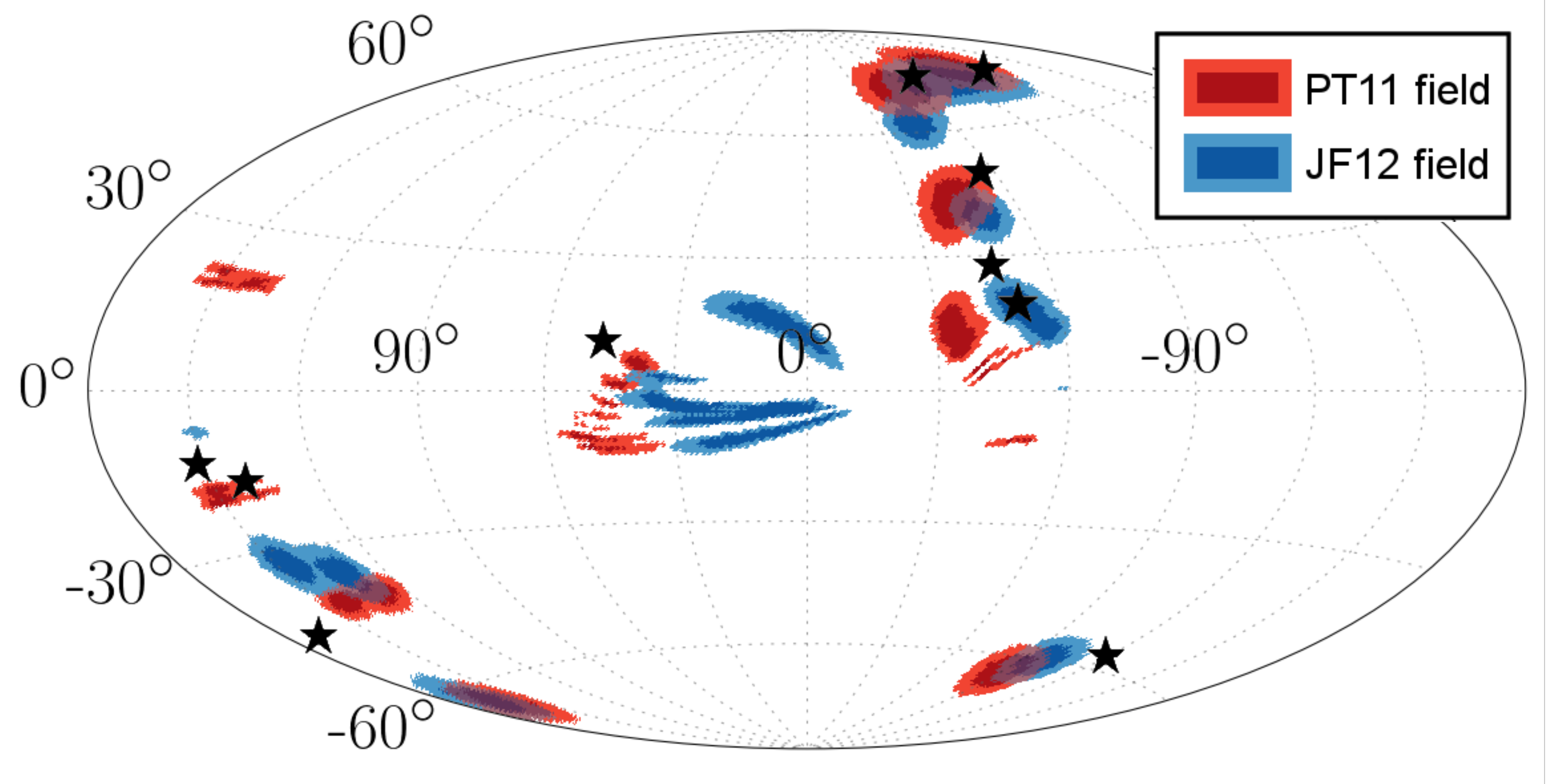}{a)}
\includegraphics[width=0.44\textwidth]{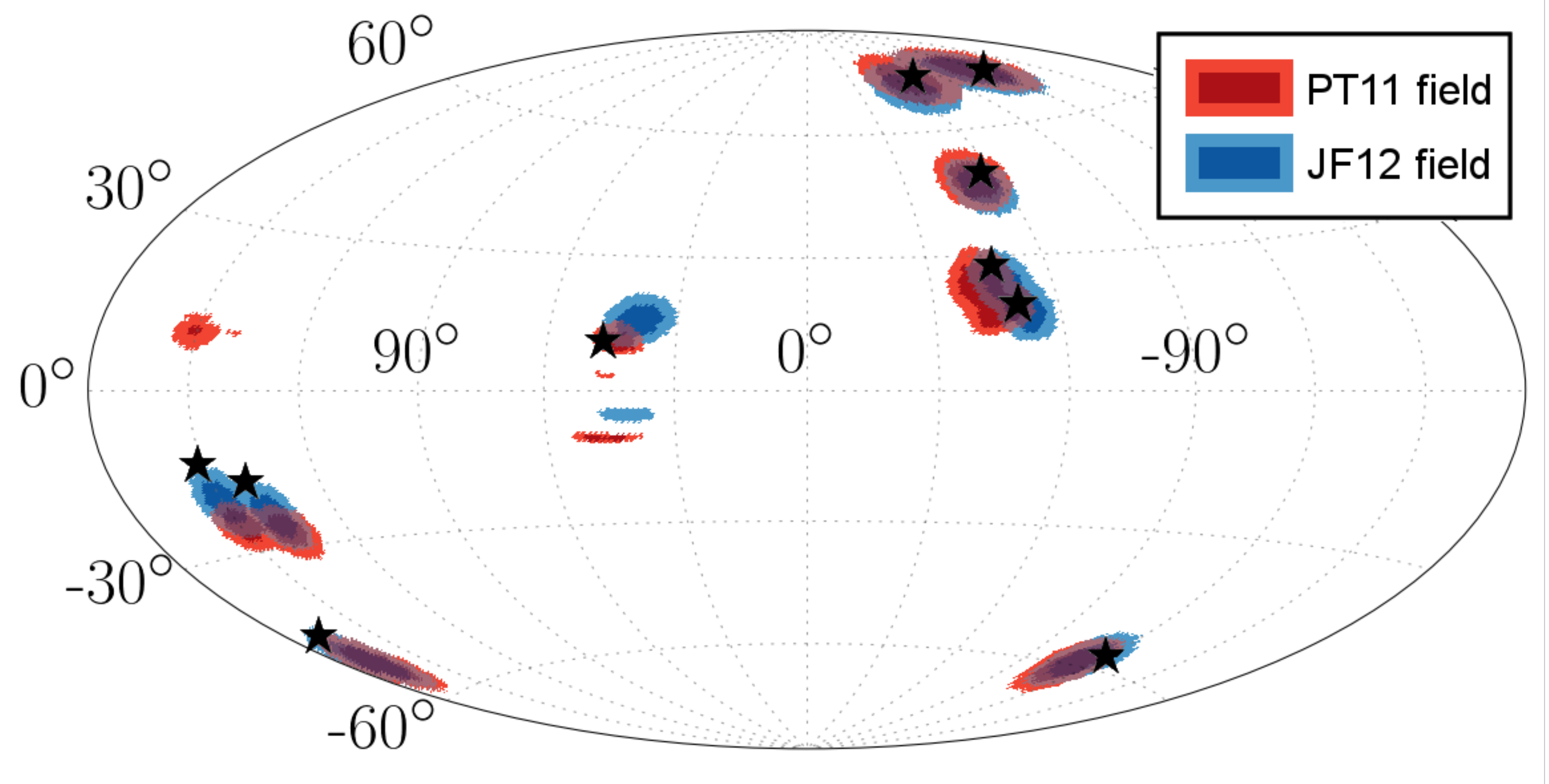}{b)}
\caption{Probability density functions reflecting 
arrival distributions of cosmic rays after traversing the PT11 
galactic magnetic field (red contours) or the JF12 field (blue contours), respectively.
The contours denote $68\%$ and $95\%$ levels.
The incoming cosmic ray distributions were centered at the directions denoted by the
star symbols and Fisher distributed with a Gaussian width of $3$~deg; rigidity 
a) $R=20$~EV, b) $R=60$~EV.
}
\label{fig:source_arrivals}
\end{figure}

In Fig.~\ref{fig:source_arrivals} we show example arrival distributions of cosmic rays
with rigidity a)~$R=20$~EV, b)~$R=60$~EV originating from ten sources.
The directions of the sources are denoted by the star symbols.
The initial cosmic ray distributions followed the Fisher distribution
with a Gaussian width of $3$~deg.

Indicated by the dark (light) red regions are the $68\%$ ($95\%$) arrival probability 
distributions of the cosmic ray after deflections by the PT11 field.
The blue regions give the corresponding arrival probability distributions from the JF12 field.

At cosmic ray rigidity $R=20$~EV (Fig.~\ref{fig:source_arrivals}a) at least half of the 
arrival probability distributions exhibit substantial overlap for the PT11 and JF12 fields.
With increasing rigidity (Fig.~\ref{fig:source_arrivals}b, $R=60$~EV) 
the overlap increases as expected.
The number of images is also reduced at larger rigidity.

This implies that, at large rigidities, the agreement of the two field 
parameterizations is sufficiently large to investigate the impact of 
the field uncertainties on a point source search which we present in the
following section.

\subsection{Simulated point source search}

For our simulated search for origins of cosmic rays we study separately sources in the three regions 
of the galaxy (latitude $\pm 19.5$~deg).
In each region we repeatedly simulate ten sources and demand the cosmic rays to follow a Fisher 
probability distribution with a Gaussian width of $3$~deg.

We also simulate isotropically distributed cosmic rays with full sky coverage as a 
background contribution.
In the following we perform multiple analyses with sets of $500$ cosmic rays for which 
we vary the contribution of signal cosmic rays, i.e. cosmic rays arriving from 
the ten sources, between signal fraction $f_s=0\%$ and $f_s=100\%$.

To quantify the analysis sensitivity we use the log-likelihood function 
\begin{equation}
\ln{L}(a) = \sum_{i=1}^{N} \; 
   \ln{\left[ \, a \, P(R_i,l_i^\prime,b_i^\prime)  + ( 1 - a ) \, B \, \right]} \;.
\label{eq:loglikeli}
\end{equation}
The sum refers to all simulated $N=500$ cosmic rays.
Parameter $a$ denotes the anticipated fraction of signal cosmic rays from the sources
when analyzing the data, 
and the isotropically distributed cosmic rays are assumed to contribute with $(1-a)$
correspondingly.
The probabilities $P(R,l^\prime,b^\prime)$ represent the anticipated arrival probability 
distributions for cosmic rays with rigidity $R$ which originate from the sources and
are expected to be observed in directions ($l^\prime,b^\prime$) on Earth.
They were obtained using the lensing techniques.
The background probability $B$ corresponds to the inverse number of pixels  
for which we used the above $N_{pix}=49,152$ pixels of approximately $1\;\deg$.

As the test statistics we use the likelihood ratio
\begin{equation}
t = 2\; \ln{\frac{L(a)}{L(a=0)}}
\label{eq:teststatistic}
\end{equation}
which approximately follows a $\chi^2$ distribution with $1$ degree of freedom \citep{Wilks1938}.
For each anticipated signal fraction $a$ we repeat the simulation of cosmic ray sets $1000$ times
and determine the average maximum $t_{max}$.
The significance by which isotropic arrival distributions can be excluded is 
then estimated by converting the integral $\int_{t_{max}}^\infty \chi^2 dt$ 
above $t_{max}$ to Gaussian standard deviations $\sigma$.

In the analysis we use as the simulated scenario the JF12 arrival probability 
distributions $P(R,l^\prime,b^\prime)$ to describe cosmic ray deflections.
To obtain a benchmark for a best-case scenario, where the field and the cosmic
ray rigidities are perfectly known, we first analyze cosmic rays 
with rigidity $R=20$~EV by using the JF12 field, here representing the true field.
Note that this scenario returns optimistic results as we neglect deflections in the 
small-scale random field and demand sources to be located in one of the three galactic 
regions exclusively.

\begin{figure}[hbt!]
\includegraphics[width=0.45\textwidth]{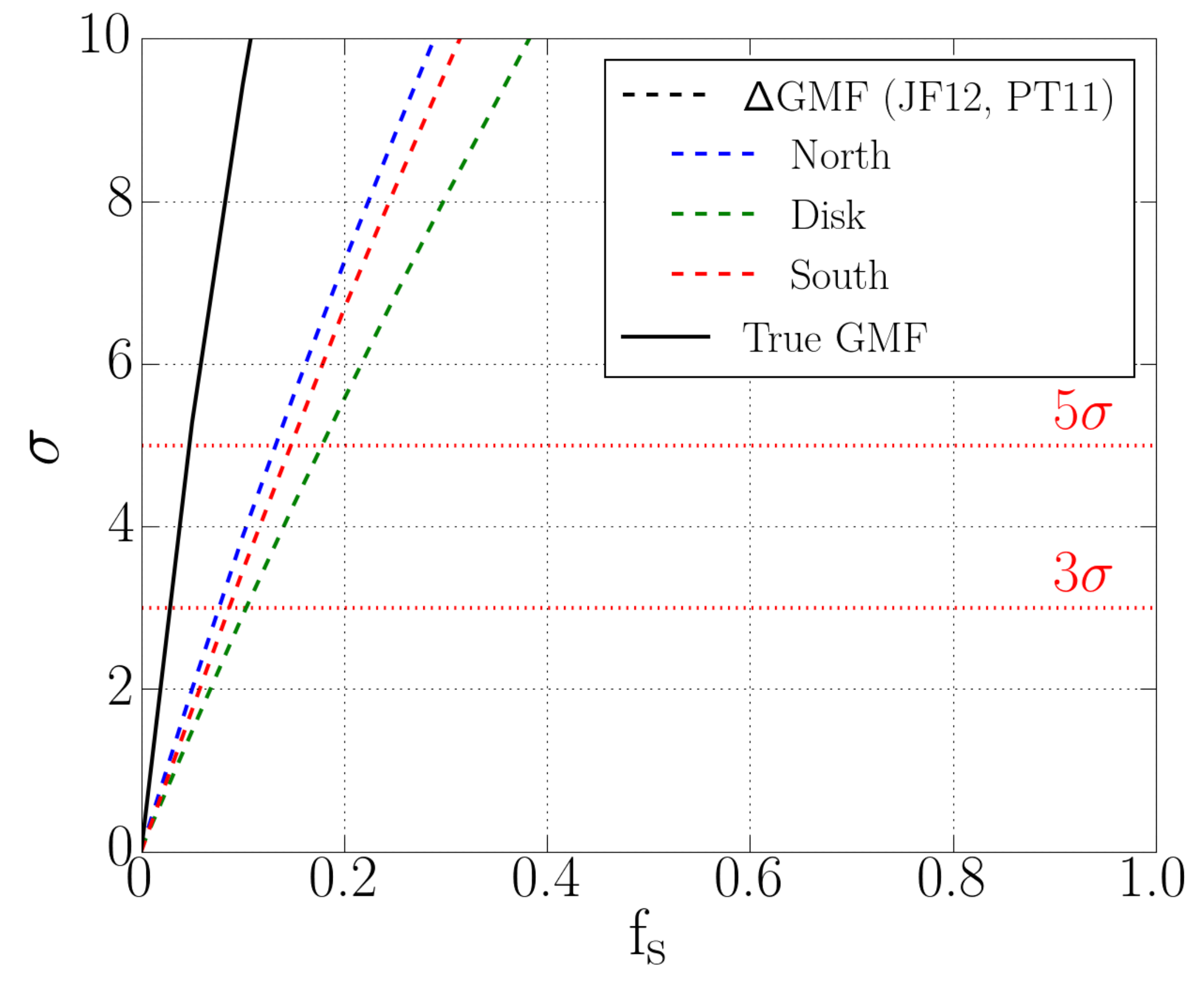}{a)}
\includegraphics[width=0.45\textwidth]{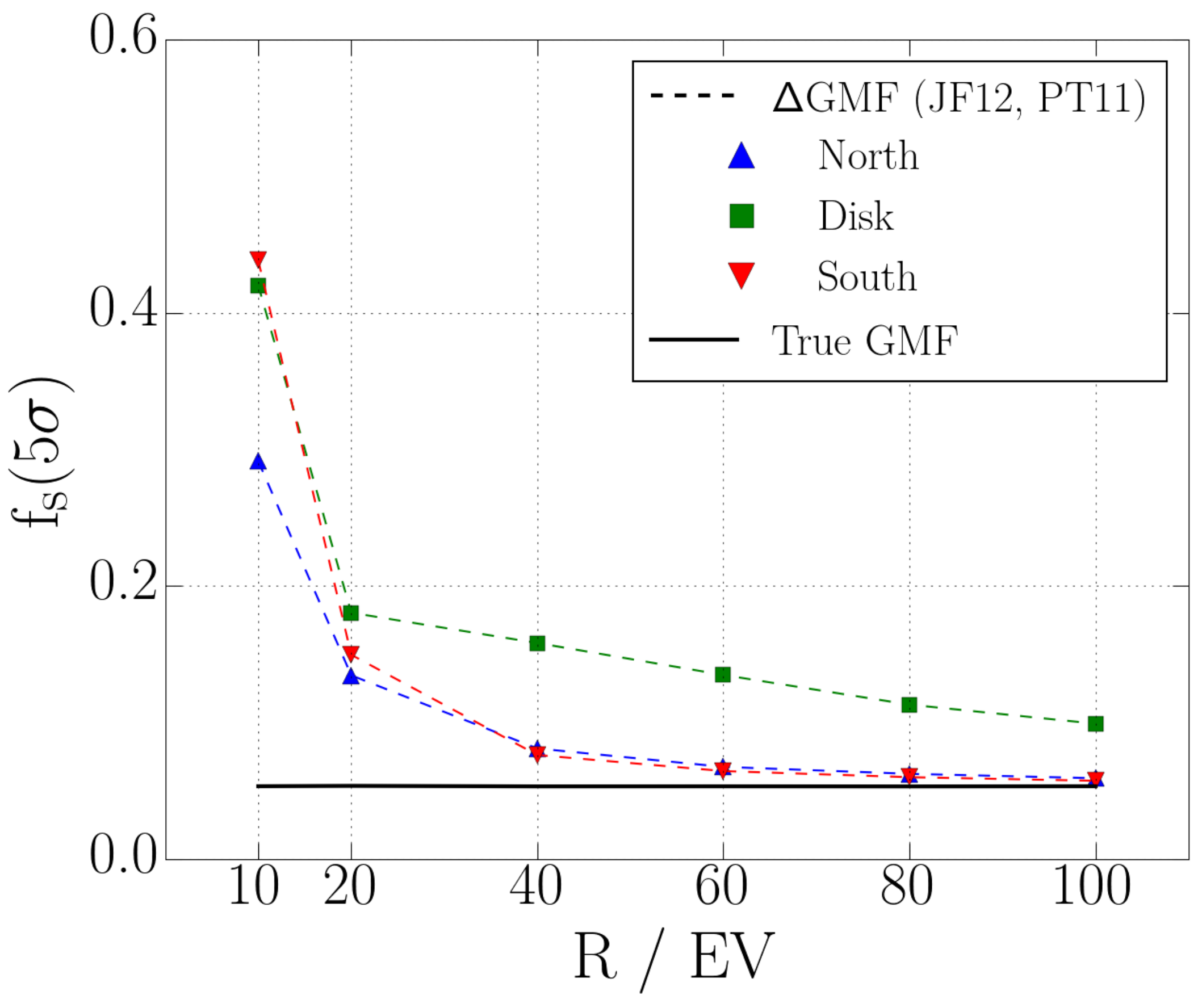}{b)}
\caption{Simulated point source search with ten sources located in one of the 
three regions separated by galactic latitudes $\pm 19.5$~deg 
and sets of $500$ cosmic rays with fraction $f_s$ originating from the sources 
and $(1-f_s)$ from isotropic background with full sky coverage.
a) Significances for deviation from isotropic arrival distributions at $R=20$~EV 
using the true galactic field (full curve) for the anticipated arrival probability 
$P$ in eq. \protect (\ref{eq:loglikeli}), and the 
other field alternatively (dashed curves).
b) Required signal fraction for a $5\sigma$ deviation from isotropic
arrival directions using the true field (full curve) compared to using
the other field (dashed curves) as a function of rigidity $R$.
}
\label{fig:sensitivity}
\end{figure}

In Fig.\ref{fig:sensitivity}a we show the significance $\sigma$ as a function
of the signal fraction $f_s$ of the simulated sample.
With perfect knowledge of the galactic field a signal fraction of $f_s=5\%$ is 
sufficient for a $5 \sigma$ discovery (full curves).

To take into account uncertainties in the galactic field as encoded in the 
two different parameterizations, we then perform the analysis with the PT11 
arrival probability distributions instead of the true JF12 probability distributions.
As the results are slightly dependent on the exact directions of the sources, 
we repeat the analyses nine times in each region and present the average resulting values.

In Fig.\ref{fig:sensitivity}a we show the significance $\sigma$ of a deviation from isotropic 
arrival distributions as a function of the average signal fraction $f_s$ for the three regions 
(dashed curves).
A signal fraction of $f_s=14\%$ is sufficient 
for a $5\sigma$ discovery for the northern and southern regions, and slightly larger 
for the disk region ($f_s=18\%$).
When compared to the above best-case scenario, the field uncertainties require the signal fraction 
for discovery to increase substantially by a factor of $3-4$.

In Fig. \ref{fig:sensitivity}b we present the required signal fraction $f_s(5\sigma)$
for discovery as a function of cosmic ray rigidity $R$.
The full curves represent the required signal fraction for the benchmark scenario 
having perfect knowledge of the galactic magnetic field. 
The required signal fraction $f_s(5\sigma)$ when including field uncertainties are
shown by the dashed curves.

It is interesting to note that, although deflections are on average twice as large 
in the southern region compared to the northern region, their sensitivities 
are of similar value above rigidity $R=20$~EV.
In the disk region the different characteristics of the two parameterizations have a large 
impact on the point source search.
Here the required signal fraction almost doubles irrespective of rigidity.

As expected, the sensitivity of the point source search improves consistently with 
increasing cosmic ray rigidity.
Again comparing with the above best-case scenario, cosmic rays with $R=60$~EV 
entering the galaxy in the northern or southern regions require a moderate 
$20\%$ increase in the required signal fraction for discovery owing to
field uncertainties.
In contrast, incoming directions at the disk region need a $2.5$-fold larger signal 
fraction.

In view of our current knowledge of the galactic field, for analyses which aim at 
selected source directions and require corresponding galactic field corrections 
we recommend a cosmic ray rigidity of at least $R=20$~EV.
In our simulated search the required signal fraction for a $5\sigma$ discovery 
remains below $20\%$.

\subsection{Summary of the influence of field uncertainties on cosmic ray arrival}

Our direct comparisons of cosmic ray deflections in the two field parameterizations 
provide some information on the actual knowledge of the entire field map. 
For cosmic ray rigidity $R=6$~EV, the angular distance after deflections in the two fields
is within $\delta=30$~deg for $50\%$ of the directions incoming to our galaxy. 
This value is much smaller at $R=60$~EV rigidity, where the differences are below 
$\delta=5$~deg for $60\%$ of the directions incoming to our galaxy. 

To study the influence of uncertainties in the field on searches for cosmic ray 
origins, we analyzed simulated astrophysical scenarios using a log-likelihood method. 
The method includes the anticipated probability distributions for cosmic ray arrival after 
traversing the galactic field, and quantifies deviations from isotropic
arrival distributions.

We estimated the influence of field uncertainties by using one field parameterization 
in the simulation of the scenario, and by applying the other field parameterization in the 
log-likelihood analysis. 
For cosmic rays with rigidity $R=60$~EV arriving from sources within galactic latitudes
$\pm 19.5$~deg (disk region) we found that the field uncertainties increase the 
required signal fraction for a $5\;\sigma$ discovery substantially by more than a 
factor of two.
However, for sources in the northern and southern regions emitting cosmic rays with
$R=60$~EV, the field uncertainties are relatively small and increase the required signal fraction
for a $5\;\sigma$ discovery by $20\%$ only.

\section{Conclusion}

Corrections for deflections in the galactic magnetic field using the two parameterizations 
PT11 and JF12 can be meaningfully considered for cosmic ray rigidities above $R>6$~EV.
Above this rigidity, deflections can be distinguished from diffusive random walk.
This has strong implications for analyses using cosmic ray data with mixed composition. 
For protons this rigidity corresponds to energies above $E=6$~EeV. 
However, when analyzing, e.g., Neon nuclei with charge $Z=10$, meaningful corrections 
can be performed for energies above $E=60$~EeV only.

When quantifying uncertainties in the galactic field from comparisons of
the two field parameterizations PT11 and JF12, the rigidity threshold needs to
be raised substantially.
Then both fields give similar predictions for cosmic ray deflections 
in the northern and southern regions with galactic latitudes $\vert l \vert > 19.5$~deg.
In the disk region $\vert l \vert < 19.5$~deg, however, the differences in the 
predictions remain large.
Consequently, in our simulated search for cosmic ray origins the arising uncertainties are 
substantial for sources near the galactic disk,
and may be considered acceptable for sources aside the disk emitting cosmic rays 
with rigidities $R\ge 20$~EV.




\section*{Acknowledgments}
We wish to thank very much G. Farrar for fruitful discussions, and 
G. Farrar, P. Tinyakov and M. Sutherland for valuable comments on the manuscript.
This work is supported by the Ministerium f\"ur Wissenschaft und Forschung, Nordrhein-Westfalen, 
the Bundesministerium f\"ur Bildung und Forschung (BMBF), and the Helmholtz Alliance 
for Astroparticle Physics.

\end{document}